\documentclass[a4paper,12pt]{article}

\usepackage{a4wide}
\usepackage{graphicx,subfigure,hhline}
\usepackage[colorlinks=true,linkcolor=blue,citecolor=red]{hyperref}
\usepackage{xcolor}
\usepackage{amssymb,amsthm,empheq,bbold}
\usepackage{hyperref}
\usepackage[inline]{enumitem}
\usepackage[ruled,vlined,linesnumbered]{algorithm2e}
\usepackage{caption} 
\usepackage[pagewise]{lineno}%\linenumbers
\usepackage{geometry}

\theoremstyle{plain} 
\newtheorem{theorem}{Theorem} 

\newtheorem{remark}{Remark}

\definecolor{ggreen}{cmyk}{1,0,1,0}

\definecolor{ggreen}{cmyk}{1,0,1,0}

\allowdisplaybreaks

\begin{document}
	%\begin{linenumbers}
	
	\title{A {nonconservative} kinetic framework for a closed-market society subject to shock events}
	
	\author{Marco Menale$^{1,\, *}$, Ana Jacinta Soares$^{2}$, Romina Travaglini$^{2,3}$\\[1em]
		$^1${\footnotesize Department of Mathematics and Applications ``R. Caccioppoli",}\\ 
		{\footnotesize University of Naples ``Federico II",}\\ 
		{\footnotesize Via Cintia, Monte S. Angelo I-80126 Napoli, Italy} \\[-1 mm]
		{\footnotesize\url{marco.menale@unina.it} (*corresponding author)}
		\\[2mm]
		$^2${\footnotesize Centre of Mathematics, University of Minho,}      
		{\footnotesize Campus of Gualtar, 4710-057 Braga, Portugal} \\[-1mm]
		{\footnotesize\url{ajsoares@math.uminho.pt}}
		\\[2mm]
		$^3${\footnotesize Istituto Nazionale di Alta Matematica "Francesco Severi",}\\ 
		{\footnotesize Dipartimento di Scienze Matematiche, Fisiche e Informatiche, Universit\`a di Parma,}\\      
		{\footnotesize Parco Area delle Scienze, 53/A, 43124 Parma, Italy} \\[-1mm]
		{\footnotesize\url{romina.travaglini@unipr.it}}}
	
	\date{ }
	
	\maketitle
	
	\begin{abstract}
		Recently, several events 
		have shockingly impacted society, carrying tough consequences. 
		However, not all individuals are similarly affected by shock events.
		Among other factors, the consequences can vary depending on the income class. In our presented work, the approach typical of kinetic theory is used to analyze the dynamics of a closed-market society exposed to various types of shock events.
		To achieve this, we introduce non-conservative equations, 
		incorporating proliferative and destructive binary interactions as well as external actions. Specifically, the latter term 
		reproduces the
		shock events, and to accomplish this, we introduce an appropriate external force field into the kinetic framework, 
		modeled using Gaussian functions. 
			Several numerical simulations
			are presented to illustrate the behavior of the solution predicted by the model and an application in comparison to real data relative to the Hurricane Katrina catastrophe is carried out.
	\end{abstract}
	
	\smallskip
	
	
	\smallskip
	
	
%%%%%%%%%%%%%%%%
%%%%%%%%%%%%%%%%
	
	\section{Introduction}
	\label{secintro}
	
	During the last few decades, the kinetic theory framework has been widely used by the 
	mathematical community for model{ling} the evolution of socio-economic systems
	\cite{bertotti2011microscopic, letizia2016economic, cordier2005kinetic, during2018kinetic, maldarella2012kinetic}. 
	Those mode{ls p}rovide a complete description of long-term outcomes in a society 
		in which the dynamics do not depend on environmental processes or other external {factors}. 
		{Moreover, the} society, or a certain portion of it, may occasionally experience some \textit{shock events} 
	with profound impact on the population, particularly on the income distribution.
	These shock events are intended, in socio-economic terms, as unforeseen incidents that have a significant
	and unpredictable influence on a large scale. 
	Socio-economists have always exerted significant efforts to understand the impact of these events on the population, 
	aiming to predict potential outcomes of crisis, see e.g.\,\cite{buheji2019framework}
	% \cite{Framework for Mitigating Coming Socioeconomic Crisis Mohamed Buheji*Dunya Ahmed}. 
	Furthermore, shocks affect distinct income classes with varied consequences. 
	For instance, the recent COVID-19 pandemic also had strong effects concerning the wealth distribution in society, as reported in \cite{narayan2022covid}.
	% \cite{COVID-19 and Economic Inequality: Short-Term Impacts with Long-Term Consequences}.
	These effects have been modeled by several applied mathematicians, see e.g.\,\cite{dimarco2020wealth},
	where an economic model has been developed in the presence of an epidemic phenomenon,
	with a focus on the social and economic impact of the epidemic on different social classes.
	Similarly, one may consider the impact of other shock events, as
	wars \cite{justino2012war} or natural disasters such as earthquakes \cite{daniell2014socioeconomic} or hurricanes \cite{shaughnessy2010income},
	% \cite{War and Poverty Patricia Justino}, 
	% \cite{Chapter 9 - The Socioeconomic Impact of Earthquake Disasters}, 
	%climate changes \cite{leichenko2014climate}
	% \cite{Climate change and poverty:vulnerability impacts and alleviation strategies, Robin Leichenko1∗and Julie A. Silva}, 
	and so forth.
	
	Motivated by this, in this paper, we aim to develop a kinetic model that describes the effects of shock events on a system in which individuals are categorized by their wealth state. It is worth pointing out that the use of kinetic tools for modeling economic phenomena has been of wide interest in the community over the decades. Among others, we refer to papers \cite{cordier2005kinetic,during2008kinetic, pareschi2014wealth, torregrossa2018fokker,bisi2009kinetic} that provide some issues about the modeling of wealth distribution and related features, for instance by using Fokker-Planck equations. Moreover, we still refer to the paper \cite{dolfin2017modeling}, along with references therein.
	
	The kinetic theory allows us to model an interacting system composed of \textit{particles}, also called \textit{agents}, that stochastically interact. These interactions determine the evolution of the overall system.
	Papers \cite{bellomo2008modeling,marsan2016stochastic} are referenced to the interested reader and provide references for additional information. Basically, it is possible to individuate three different levels for describing the system, namely a \textit{microscopic} a \textit{mesoscopic}, and a \textit{macroscopic} one.
	The microscopic level describes the binary stochastic interactions among pairs of particles. 
	More precisely, the outcomes of such interactions are regulated by means of a real variable that can be continuous or discrete. 
	The mesoscopic
	level represents the statistical level since the interest is focused on the distribution of particles.  At this level,
	suitable equations depict the evolution of the system; in particular, they can be 
	either integro-differential equations or ordinary differential equations, depending on the nature of the
	microscopic variable,  \cite{bertotti2011microscopic, cordier2005kinetic, herrmann2012self}. Finally, at the macroscopic scale, the evolution of the whole system, as a single entity, is derived as, for instance, the hydrodynamic limit of the mesoscopic equations, \cite{pareschi2019hydrodynamic}.
	
	The classical kinetic approach regards the evolution of conservative systems, 
	i.e. systems {that are not} subjected {to} external actions {or} nonconservative 
	(proliferative/destructive) binary interactions where the total amount of particles {or} agents is conserved. 
	{However, excluding} these aspects would be too restrictive to {provide a} realistic 
	{representation} of the evolution of some phenomena. 
	Recently, some {works have been} developed in this direction. 
	Among others, in \cite{carbonaro2023nonconservative}, the external action is modeled by introducing a suitable external force field.
	{Th}us, the evolution is analyzed with respect to the specific choice of the external force field. 
	{In} \cite{menale2023kinetic}, {the} authors discuss the analytical properties of a kinetic system with 
	time-dependent nonconservative parameters, i.e. birth {and} death events. 
	{Bo}th papers {apply these} models {to} ecological systems, 
	{usi}ng a predator-prey approach.
	
This paper aims to develop this modeling for the particular socio-economic framework in which a subgroup of the society is affected by a shock event. The nonconservative external force field is adopted to describe the migration of individuals away from the considered affected fraction, while birth and death events describe physiological oscillations in the population. First, a closed-market society is assumed, where individuals interact solely through money exchange, for details, see \cite{letizia2016economic, chakraborti2000statistical, chatterjee2007kinetic}, and references therein. The inclusion of nonconservative terms provides a more realistic setting for modeling the evolution of socio-economic systems. In particular, in the spirit of \cite{menale2023kinetic}, time-dependent nonconservative binary interactions are assumed. Moreover, an external force field with a specific analytical shape acts on the system which can resemble the impact of shock events. Additionally, diverse external force fields are considered, as shock events can be characterized by different features, such as intensity or time occurrence. In this context, we discuss and compare the related scenarios with each other.
	
	 The model here proposed is then applied to the specific event of Hurricane Katrina, which impacted the city of New Orleans in 2005. In this case, the external force represents the hurricane determining a change in the population, since inhabitants from different income classes moved to another city after the event and a part of them did not return \cite{shaughnessy2010income}. This can be appropriately interpreted as a nonconservative phenomenon.
	
	From the analytical point of view, the introduction of nonconservative terms in the kinetic framework, such as nonconservative binary interactions and external force field, may result in the loss of some properties. For instance, global existence and positivity of solutions are not generally ensured. Indeed, in \cite{arlotti1996qualitative}, authors show that blow-up phenomena in finite time may occur. Therefore, the analysis of such a kinetic framework is a delicate procedure, and some restrictions are mandatory on nonconservative parameters themselves, i.e. on the shape of both nonconservative binary interactions and external force field. In this direction, some first results emerged in the previous papers \cite{carbonaro2023nonconservative, menale2023kinetic}. 
	
	It is worth mentioning that the analysis of the impact of a shock event on a segment of a society is particularly relevant for replicating the dynamics of diverse income classes right after the shock or in a short-time period. For the long-time behavior, it is reasonable to assume that the shock event does not affect the dynamics, which can be described by the already existing conservative models, or those endowed with only birth and death nonconservative processes. Nevertheless, the long-term configuration of the wealth distribution in the society may differ from the scenario where no shock occurs. 
	
	We provide here an overview of the organization of this paper. In Section \ref{secmodel}, we present and discuss the nonconservative kinetic framework, with time-dependent nonconservative binary interactions and a specific external force field. In particular, we prove a theorem of existence and uniqueness of local solution and discuss more general scenarios. Section \ref{secmark} is devoted to applying the kinetic nonconservative framework to model a closed-market population affected by a shock event. 
	%In Subsection \ref{sseckin} and Subsection \ref{ssecks}, the framework is specialized in two situations, i.e. for $3$ and $5$ income classes. Moreover, the
	In particular, the analytical shape of the external force field describing the shock events is provided. Subsection \ref{ssecnum} includes numerical simulations showing the short-time and long-time behavior of solutions in comparison to different scenarios.
	Then, the model is adapted in Section \ref{secS5} to real data on the impact of Hurricane Katrina on the population of New Orleans, with further simulations performed for the specific case. Finally, Section \ref{secconcl} presents some conclusions and future research perspectives. 
	
	%%%%%%%%%%%%%%%%	
	
	\section{Kinetic framework}
	\label{secmodel}
	
	In a classical discrete kinetic theory framework \cite{kawashima1987large,bellomo1991discrete,kawashima1991global,monaco1994chapman,jin1998diffusive,oliveira2005global}, 
	the overall system is segmented into \textit{functional subsystems}, where agents within each subsystem adhere to a common strategy. The interpretation of such a strategy varies depending on the specific application under consideration.
	
	The \textit{microscopic state} is characterized by the real discrete variable $u$, generally referred as \textit{activity}, ranging in a 
	finite subset of $\mathbb{R}$, that is
	$$
	u \in \{u_1, u_2, \dots, u_n\}\subseteq\mathbb{R}.
	$$ 
	For the aim of the present work, as well as to avoid overcomplicating the mathematics in this work, we suppose that agents within the $i$-th functional subsystem share the same microscopic state $u_i$. Roughly speaking, we establish a correspondence between the microscopic variable and functional subsystem.
	
	The agents may undergo \textit{binary stochastic interactions} described by the following quantities: 
	\begin{itemize}
		\item The \textit{interaction rate} $\eta_{hk}\geq 0$, $h,k \in \{1, 2, \dots, n\}$, that furnishes the number of encounters {per unit of time} between agents of the $h$-th and $k$-th functional subsystems.
		
		\item The \textit{transition probability} $B^i_{hk}$,  $i,h,k \in \{1, 2, \dots, n\}$, that provides the probability of an agent from the $h$-th functional subsystem transitioning into the $i$-th subsystem, after an interaction with an agent from the $k$-th {functional subsystem}; being $B^i_{hk}$ a probability, we remark that it holds
		\begin{equation}
			\label{assump}
			\sum_{i=1}^nB^i_{hk}=1, \qquad \forall h,k \in {\{1, 2, \dots, n\}.}
		\end{equation}
		
		\item The \textit{proliferative/destructive rate}, i.e. the function
		$$
		{\mu_{hk}} :  [0, \,T]\rightarrow \mathbb{R},\qquad T>0,
		$$
		with $h,k \in \{1,2, \dots, n\}$, that amounts the nonconservative events due to interactions between 
		agents of the $h$-th and {$k$-th functional subsystems}. 
	\end{itemize}
	In our model, we assume time independence for interaction rates and transition probabilities, whereas proliferative and destructive rates are assumed to be time-dependent. This choice is related to the specific aims of this paper. 
	In particular, the latter parameters $\mu_{hk}(t)$ contribute to the nonconservative dynamics of the system. 
	
	At the \textit{mesoscopic level}, for $i \in \{1, 2, \dots, n\}$, we introduce the \textit{distribution function}
		$$
		f(t,u_i) : [0,\, T]\rightarrow\mathbb{R}^+, \qquad T>0,$$
		such that the quantity $f(t,u_i)$ represents the number of agents at time $t>0$ in the microscopic state $u_i$ or, equivalently to what has been stated above, the number of agents at time $t>0$ in the $i$-th functional subsystem. For the sake of simplicity, we use the notation $f_i(t)$, for $t\geq0$, instead of  $f(t,u_i)$.
		In a more general context, the distribution function writes $f_i(t,u_j)$, and it refers to the amount of agents in the $i$-th functional subsystem, with microscopic state $u_j$.
	
	Furthermore, we introduce the \textit{vector distribution function} of the overall system as follows:
	$$
	\mathbf{f}(t)=\left(f_1(t), f_2(t), \dots, f_n(t)\right).
	$$
	We make the assumption that the external environment influences the time evolution of the system, as recently done in \cite{carbonaro2023nonconservative}. Therefore, an \textit{external force field} is introduced by means of the following functional:
	$$
	\mathbf{F}[\mathbf{f}](t)  : [0,\, T]\rightarrow \mathbb{R}^n.
	$$
	Then, for $i \in \{1, 2, \dots, n\}$, the component $F_i[\mathbf{f}](t)$ represents
	the action of this external field over the $i$-th functional subsystem. 
	
	The field $\mathbf{F}[\mathbf{f}](t)$ may assume different shapes 
	depending on the particular application being considered. 
	For the aims of the present study, the following choice is provided:
	\begin{equation}
		\label{forcshape}
		F_i[\mathbf{f}](t)=\lambda_i(t)\,f_i(t), \qquad \forall i \in \{1, 2, \dots, n\}.
	\end{equation}
	Then, we may observe that
	\begin{equation}
		\label{eqf}
		f_i(t)=0 \Rightarrow F_i[\mathbf{f}](t)=0.
	\end{equation}
	
	Generally speaking, the function $\lambda_i(t)$ represents the action of the external force field over the $i$-th functional subsystem 
	independently of its current state; nevertheless, this action is weighted by the state of the $i$-th functional subsystem at time $t\geq 0$. %Therefore, the impact of $F_i[\mathbf{f}](t)$, for $i \in \{1, 2, \dots, n\}$, depends on the distribution function $f_i(t)$, as given in \eqref{forcshape}.   
	
	Accounting the binary and stochastic microscopic interactions, the \textit{kinetic equation} of the $i$-th functional subsystem, for 
	$i \in \{1, 2, \dots, n\}$, writes
	\begin{equation}
		\label{eqkin}
		\frac{df_i}{dt}(t)=\sum_{h,k=1}^{n}\eta_{hk}B^i_{hk}f_h(t)f_k(t)-f_i(t)\sum_{k=1}^{n}\eta_{ik}f_k(t)
		+ f_i(t)\sum_{k=1}^{n}\eta_{ik}\mu_{ik}(t)f_k(t) + \lambda_i(t)f_i(t).
	\end{equation}
	The first two terms on the right-hand side account for the conservative dynamics as it is classically depicted in literature (see e.g. \cite{bertotti2010modelling}). The third and the fourth ones, instead, represent the nonconservative process. More specifically, the former describes the proliferative/destructive interactions among individuals, while the latter stands for the impact of the force field on the $i$-th functional subsystem. More details about the role of these terms will be provided in the next sections from the viewpoint of the application of the model to real phenomena. 
	Equation \eqref{eqkin} is a first-order nonlinear ordinary differential equation,
	exhibiting quadratic nonlinearities. By straightforward computations, the kinetic equation \eqref{eqkin} rewrites
	\begin{equation}
		\label{eqkin2}
		\frac{df_i}{dt}(t) \!=\!\! \sum_{h,k=1}^{n} \! \eta_{hk}B^i_{hk}f_h(t) \! f_k(t) \!-\! f_i(t) \!
		\sum_{k=1}^{n} \! \eta_{ik}f_k(t) \!+\! f_i(t) \! \left( \! \sum_{k=1}^{n} \! \eta_{ik}\mu_{ik}(t)f_k(t) \!+\! \lambda_i(t) \! \right) \!,
	\end{equation}
	in such a way that conservative and nonconservative dynamics are well separated and the direct dependence of the latter 
	dynamics on the $i$-th distribution function is more evident.
	
	\medskip
	
	The \textit{$p$th-order moments} are defined as
	$$
	\mathbb{E}_p[\mathbf{f}](t):=\sum_{i=1}^nu_i^p\,f_i(t),
	$$
	for $p \in \mathbb{N}$, and describe the \textit{macroscopic state} of the system.
	From probabilistic and statistical standpoint, the $0$th-order moment represents the density, the $1$st-order moment stands for the mean, and the $2$nd-order moment is related to the variance. However, from a physical perspective, the $0$th, $1$st, and $2$nd order moments can be interpreted as density, linear momentum, and kinetic energy, respectively, for the entire system. For socio-economic applications, these moments have a particular interpretation, due to the particular dynamics of these phenomena. The $0$th-order moment, also indicated by $\rho(t)$, represents the number of agents in the overall system at time $t>0$. Moreover, if $\rho(t)\neq 0$, the overall system average, $m(t)$, or the mean value with respect to the activity variable, 
	is provided by the $1$st-order moment, i.e. (see \cite{della2023intransigent} for details)
	$$m(t):=\frac{1}{\rho(t)}\mathbb{E}_1[\mathbf{f}](t)=\frac{1}{\rho(t)}\sum_{i=1}^{n}u_i\,f_i(t).$$
	The dynamics of the total density 
	$\rho(t)$, that can be recovered from the kinetic equations \eqref{eqkin2}, bearing the assumption \eqref{assump} in mind,
	is such that
	$$\dot{\rho}(t)=\sum_{i=1}^nf_i(t)\left(\sum_{k=1}^{n}\eta_{ik}\mu_{ik}(t)f_k(t)+\lambda_i(t)\right) ,$$
	which emphasizes the nonconservative nature of the system.

	Hereafter, the following analytical assumptions regarding the new terms are considered:
	\begin{align*}
		\mu_{hk}(t)&\in C([0,\,+\infty)),\\
		\lambda_i(t)&\in C([0,\,+\infty)),
	\end{align*}
	for $h,k,i\in \{1, 2, \dots, n\}$ and  $t\geq 0$.

	Assigning a positive initial data $\mathbf{f}^0\in \left(\mathbb{R}^+\right)^n$, 
	the \textit{Cauchy problem} associated to the kinetic framework \eqref{eqkin} can be formulated as
	\begin{equation}
		\label{cauchy}
		\begin{cases}
			\displaystyle 
			\frac{df_i}{dt}(t)=P_i[\mathbf{f}](t) ,
			\quad  t\in [0,\, T]
			\\[5mm]
			\mathbf{f}(0)=\mathbf{f}^0,
		\end{cases}
	\end{equation}
	where the operator $P_i[\mathbf{f}]$ is defined by
	\begin{equation}
		\label{Pi}
		P_i[\mathbf{f}](t):=\sum_{h,k=1}^{n}\eta_{hk}B^i_{hk}f_h(t)f_k(t)-f_i(t)\sum_{k=1}^{n}\eta_{ik}f_k(t)
		+f_i(t)\sum_{k=1}^{n}\eta_{ik}\mu_{ik}(t)f_k(t)+\lambda_i(t)f_i(t),
	\end{equation}
	and the following analytical result holds.
	\begin{theorem}\label{thm1}
		Let us consider the Cauchy problem \eqref{cauchy}. Let the following assumptions hold:
		\begin{itemize}
			\item there exists $\eta>0$ such that
			\begin{equation}\label{assump2}
				\eta_{hk}\leq \eta, \qquad \forall h,k \in \{1, 2, \dots, n\};
			\end{equation}
			
			\item there exists $\mu>0$ such that		
			\begin{equation}\label{assump3}
				\mu_{hk}(t)\leq \mu, \qquad \forall h,k \in \{1, 2, \dots, n\},\, \forall t>0;
			\end{equation}
			
			\item there exists $\lambda > 0$ such that
			
			\begin{equation}\label{assump4}
				|\lambda_i(t)|\leq \lambda, \qquad \forall i \in\{1, 2, \dots, n\},\, \forall t>0;
			\end{equation}
			
			\item the {assigned initial data} is such that 		
			\begin{equation}\label{assump5}
				\mathbf{f}^0\in \left(\mathbb{R}^+\right)^n.
			\end{equation}
		\end{itemize}
		Moreover, let the assumption \eqref{assump} hold.
		Then, there exists a unique bounded and nonnegative function $\mathbf{f}(t)$, 
		local solution of the Cauchy problem \eqref{cauchy} in the time interval $[0,\, t_0]$, for $t_0>0$.
		
		\begin{proof}
			For any $i \in \{1, 2, \dots, n\}$, the operator $P_i[\mathbf{f}]$
			defined in \eqref{Pi} maps the space $C([0,\, T])$ into itself. Moreover, let be $\mathbf{f}(t),\, \mathbf{g}(t) \in \left(C([0,T])\right)^n$. 
			Thus, it is possible to find a constant $c>0$ such that
			$$
			\|\mathbf{f}(t)\|_{\left(C([0,T])\right)^n},\,\|\mathbf{g}(t)\|_{\left(C([0,T])\right)^n}\leq c, \qquad \forall t>0.
			$$
			The operator $P_i[\mathbf{f}]$,
			introduced in \eqref{Pi}, can be written as
			$$
			P_i[\mathbf{f}] := G_i[\mathbf{f}] - L_i[\mathbf{f}] + N_i[\mathbf{f}] + F_i[\mathbf{f}],
			$$	
			where
			\begin{align*}
				G_i[\mathbf{f}](t)&:=\sum_{h,k=1}^{n}\eta_{hk}B^i_{hk}f_h(t)f_k(t),\\
				L_i[\mathbf{f}](t)&:=f_i(t)\sum_{k=1}^{n}\eta_{ik}f_k(t),\\
				N_i[\mathbf{f}](t)&:=f_i(t)\sum_{k=1}^{n}\eta_{ik}\mu_{ik}(t)f_k(t),\\
				F_i[\mathbf{f}](t)&:=\lambda_i(t)f_i(t).
			\end{align*}
			By using assumptions \eqref{assump}, \eqref{assump2}, \eqref{assump3}, \eqref{assump4}, 
			some standard computational arguments, as those ones developed in \cite{bertotti2011microscopic, arlotti1996qualitative}, allow to conclude that
			\begin{equation}
				\label{eqth12}
				\begin{split}
					\sum_{i=1}^{n}\left| G_i[\mathbf{f}](t)-G_i[\mathbf{g}](t) \right|
					& \leq \eta \sum_{h=1}^{n}|f_h|\sum_{k=1}^{n}|f_k-g_k|+\eta \sum_{k=1}^{n}|g_k|\sum_{h=1}^{n}|f_h-g_h|\\
					& \leq 2\eta c \sum_{i=1}^{n}|f_i-g_i|,
				\end{split}
			\end{equation}
			\begin{equation}\label{eqth13}\begin{split}
					\sum_{i=1}^{n}\left|L_i[\mathbf{f}](t)-L_i[\mathbf{g}](t)\right|&\leq\eta\sum_{i=1}^n|f_i|\sum_{k=1}^n|f_k-g_k|+\eta\sum_{i=1}^n |f_i-g_i|\sum_{k=1}^n|g_k|\\&
					\leq 2\eta c\sum_{i=1}^{n}|f_i-g_i|,
				\end{split}
			\end{equation}
			\begin{equation}\label{eqth14}\begin{split}
					\sum_{i=1}^{n}\left|N_i[\mathbf{f}](t)-N_i[\mathbf{g}](t)\right|&\leq \eta \mu \sum_{i=1}^{n}|f_i|\sum_{k=1}^{n}|f_k-g_k|+\eta \mu \sum_{i=1}^{n}|f_i-g_i|\sum_{k=1}^n|g_k|\\&
					\leq 2\eta \mu c \sum_{i=1}^{n}|f_i-g_i| ,
			\end{split}\end{equation}
			and 
			\begin{equation}\label{eqth15}\begin{split}
					\sum_{i=1}^{n}\left|F_i[\mathbf{f}](t)-F_i[\mathbf{g}](t)\right|&\leq \sum_{i=1}^{n}\left|\lambda_i(t)f_i(t)-\lambda_i(t)g_i(t)\right|\\&
					\leq \lambda \sum_{i=1}^{n}|f_i(t)-g_i(t)|.
				\end{split}
			\end{equation}
			Therefore, by using \eqref{eqth12}, \eqref{eqth13}, \eqref{eqth14} and \eqref{eqth15}, one has
			\begin{equation}
				\label{eqth16}
				\sum_{i=1}^{n}\left|P_i[\mathbf{f}](t)-P_i[\mathbf{g}](t)\right|
				\leq \left(4\eta c+ 2\eta \mu c +\lambda\right)\sum_{i=1}^{n}|f_i(t)-g_i(t)| ,
			\end{equation}
			and by \eqref{eqth16}, it follows
			\begin{equation}
				\label{eqth17}
				\left\|\mathbf{P}[\mathbf{f}](t)-\mathbf{P}[\mathbf{g}](t)\right\|_{\left(C([0,T])\right)^n}
				\leq C\|\mathbf{f}(t)-\mathbf{g}(t)\|_{\left(C([0,T])\right)^n},
			\end{equation}
			and the constant $C > 0$ is determined by both the initial data $\mathbf{f}^0$ and the system parameters.
			Condition \eqref{eqth17} proves that operator $\mathbf{P}[\mathbf{f}](t)$ is Lipschitz. 
			Therefore, standard analytical results \cite{coddington2012introduction} ensure existence and uniqueness of a local solution 
			$\mathbf{f}(t)=\left(f_1(t), f_2(t), \dots, f_n(t)\right)$ of the Cauchy problem \eqref{cauchy}, 
			in the time interval $[0,\, t_0]$, where $t_0>0$ depends on the initial data $\mathbf{f}^0$ and on the system parameters.
			
			\medskip
			
			To conclude the proof of this Theorem, the nonnegativity of the solution $\mathbf{f}(t)$ has to be proved. To this aim, we make use of the compact form \eqref{eqkin2} of the kinetic system, rearranging the terms, for $i \in \{1, 2, \dots, n\}$, as follows
			\begin{equation*}
				\frac{df_i}{dt}(t)=\left(\sum_{h,k=1}^{n}\eta_{hk}B^i_{hk}f_h(t)f_k(t)\right)+f_i(t)\left(\sum_{k=1}^{n}\eta_{ik}f_k(t)\left(\mu_{ik}-1\right)+\lambda_i(t)\right),
			\end{equation*}
			or, equivalently,
			\begin{equation}\label{eqth18}
				\frac{df_i}{dt}(t)=G_i[\mathbf{f}] {(t)}+ f_i(t)Q_i[\mathbf{f}](t),
			\end{equation}
			where
			\begin{align*}
				Q_i[\mathbf{f}](t)&:=\sum_{k=1}^{n}\eta_{ik}f_k(t)\left(\mu_{ik}(t)-1\right)+\lambda_i(t).
			\end{align*}
			Now, let, for $i \in \{1, 2, \dots, n\}$,
			$$\gamma_i(t):=\int_0^t Q_i[\mathbf{f}](s)\, ds.$$
			Then, the equation \eqref{eqth18} can be recast as, for $i \in \{1, 2, \dots, n\}$,
			
			\begin{equation}\label{eqth19}
				f_i(t)=f_i^0\exp\left(\gamma_i(t)\right)+\int_0^t{G_i[\mathbf{f}](s)}\exp\left(\gamma_i(t)-\gamma_i(s)\right)\, ds.
			\end{equation}
			The assumption \eqref{assump5}, together with the nonnegativity of the operator $G_i[\mathbf{f}](t)$ and the properties of the exponential function, ensure the nonnegativity of locally in time solution $\mathbf{f}(t)$. The proof is then complete.
		\end{proof}	
	\end{theorem}
	
	\noindent 
	Generally speaking, Theorem \ref{thm1} does not allow to ensure, for the Cauchy problem \eqref{cauchy}, the global-in-time existence of a positive and bounded solution. Indeed, when nonconservative terms are considered, blow-up phenomena may occur \cite{arlotti1996qualitative}. Nevertheless, some particular analytical shapes of the nonconservative time-dependent rates $\mu_{hk}(t)$, for $h,k \in \{1, 2,\dots, n\}$, and/or of the external force field $F_i[\mathbf{f}](t)=\lambda_i(t)f_i(t)$, for $i \in \{1, 2, \dots, n\}$, might provide the boundedness for the solution $\mathbf{f}(t)$, with the consequent global existence. Furthermore, particular relations between coefficients $\mu_{hk}(t)$ and $\lambda_i(t)$, might avoid blow-up phenomena as well. 
	For instance, positive proliferative parameters might be bounded by one of the components, $\lambda_i(t)$, of the external force field. 
	{Moreover, despite being the regularity assumptions on the functions $\lambda_i(t)$, for $i \in \{1, 2, \dots, n\}$, necessary for existence and uniqueness, at least locally in time, they could be restrictive in view of particular applications, as will be shown below, as shown in the next Section.}

	%%%%%%%%%%%%%%%%
	
	\section{Society under shock events}
	\label{secmark}
	
	In this section, we employ the kinetic framework \eqref{eqkin} to model the evolution of a society within a closed-market system.
	The key feature in our model is the integration of time-dependent nonconservative rates and the impact imparted by an external force field into the description of the closed-market society. In our description, the society is considered closed with respect to money exchanges, but open in the sense that the total population is not conserved. The total number of individuals may fluctuate, due to birth and death events or change significantly after the occurrence of a shock.
	We then illustrate different scenarios by performing some numerical simulations and discuss
	some features of the considered economic system concerning diverse choices of nonconservative terms. 
	We intend to highlight the impact of both the time-dependent proliferative/destructive rates $\mu_{hk}(t)$
	and external force field $\mathbf{F}[\mathbf{f}](t)$ 
	on the evolution of the overall system, picking a particular function shape of the latter term. 
	
	%%%%%%%%%%%%%%%%
	
	\subsection{The framework}
	\label{sseckin}
	
	We take, as a starting point, a model for a closed-market society, in which individuals interact by only money exchange. At the microscopic level, this society is endowed with an activity variable that indicates the income class. Subsequently, the entire population comprising the society is split into $n$ functional subsystems. Within each subsystem, individuals share a common wealth state, quantified by their mean income. 
	%In essence, these functional subsystems serve as representations of the various income classes of society. To simplify matters, we focus solely on the wealth state as the only feature considered for modeling the evolution of the present economic system. We are aware of this limitation in an economic setting. However,
	This choice serves our purpose of investigating the role of the new feature of this paper, 
	i.e. time-dependent nonconservative rates and the action of the external force field.

	In what follows, the scenario under inquiry is specialized in two situations with $n=3$ and $n=5$ functional subsystems, respectively. 
	If the former case allows us to provide an overview of the evolution and behavior of each income class, 
	the latter depicts a more realistic situation, being the subdivision of society more structured and complex. 
	From a socio-economic viewpoint, these income classes can be seen as follows. 
	In particular, in the case $n=3$, they represent, respectively:
	\begin{itemize}
		\item lower class, {$i=1$};
		\item middle class, {$i=2$};
		\item upper class, {$i=3$}.
	\end{itemize}
	Moreover, in the case $n=5$, these classes stand for:
	\begin{itemize}
		\item lower class, {$i=1$}; 
		\item lower-middle class, {$i=2$};
		\item middle class, {$i=3$};
		\item upper-middle class, {$i=4$};
		\item upper class, {$i=5$}.
	\end{itemize}
	For each case, $f_i(t)$ represents the distribution function {of} the $i$-th functional subsystem with respect to time. 
	{Moreover, as done in \cite{bertotti2023modelling}, we define $n$ values $0=\xi_0<\xi_1,\ldots,\xi_n$, 
	such that each interval $\left[\xi_{i-1},\xi_i\right)$, for 
	{ $i=1,\ldots,n$}  $i=1,\ldots,n-1$, identifies the income range of the {$i$-th} class. 
	The intervals are supposed to have non-decreasing length. {In} other words, $\xi_{i+1}-\xi_{i}=\beta\,(\xi_{i}-\xi_{i-1})$, 
	with $\beta\geq 1$. We denote by 
		$
		r_i = \dfrac{\xi_i + \xi_{i-1}}{2}, \quad \text{for } i = 1, \ldots, n,
		$
		the average income associated with the $i$-th income class.  
		As a modeling assumption, we consider that all individuals whose income $\xi$ falls within the interval $[\xi_{i-1}, \xi_i)$ 
		are represented by the same average income $r_i$.
		Thus, the total income of the considered population at time $t$ is given by
		$$
		\overline\xi(t)=\sum_{i=1}^n f_i(t)\,r_i.
		$$   
		}
	
	%%%%%%%%%%%%%%%%
	
	\subsection{The construction of the kinetic system}
	\label{ssecks}
	
	In order to specialize the nonconservative kinetic framework \eqref{eqkin} for the current economic system,
	the parameters {must be properly specified.}  %have to be properly outlined. 
	
	{We start by {setting} the upper bound of the lower class income as $\xi_1=3.779$ 
	and we take the parameter for the growing length of each income interval {as} $\beta=1.35$.}
	{Next, we specify} the conservative parameters of the system \eqref{eqkin}, i.e. interaction rate $\eta_{hk}$ and transition probability $B^i_{hk}$, 
	{which describe} the economic binary stochastic interactions between pairs of individuals. 
	Specifically, these conservative terms represent {monetary exchanges} in society. 
	
	These coefficients are definitively {fixed}, since our main aim is the analysis of the dynamics of the system concerning the nonconservative, 
	time-dependent, terms, i.e. proliferative/destructive rates $\mu_{hk}(t)$, and the external force $\mathbf{F}[\mathbf{f}](t)$ with the analytical shape of type    
	\eqref{forcshape}. 
	Taking into account all the aforementioned considerations, the conservative parameters are chosen as follows.
	The interaction rates are assumed constant, namely
	\begin{equation}
		\eta_{hk}=1, \qquad \forall h,k \in \{1, 2, \dots, n\}.
		\label{parmetri conservativi}
	\end{equation}
	Therefore, we are considering a society in which economic interactions among individuals belonging to different functional subsystems, 
	i.e. with different wealth-state, are uniform. 
	{This assumption might appear rather restrictive. However, it aligns with the closed-market structure of the society considered here
	(see for details \cite{chakraborti2000statistical,bertotti2010modelling}). 
	Moreover, it is useful as it allows the removal of certain parameters from the kinetic system \eqref{eqkin2}, 
	thereby slightly reducing the complexity, both analytical and numerical. In the framework of the application of the model to a real scenario, we will consider a different choice for the rates.}
	
	{We report here the expression of the transition probability $B^i_{hk}$, assigned in line with {the recent framework} 
	developed in \cite{bertotti2023modelling}.
	We start by defining the baseline terms	
	$a^i_{hk} =\delta_{ih}$, with $\delta_{ih}$ being the Kroneker delta. 
	Next, we introduce the deterministic interaction probabilities $p_{hk}$, that satisfy:
	$
	p_{1,k} = 0,\, p_{h,n} = 0,
	$
	and for $h = 2, \ldots, n$, $k = 1, \ldots, n-1$,
	$$
	p_{hk} = c_d \dfrac{\min(r_h, r_k)}{r_n},
	$$
	where $c_d$ is a fixed constant (in our case $c_d=0.25$).
	Using these probabilities, the transition coefficients $c^i_{hk}$ are defined as follows. Fixing  $S$  the amount paid in a transaction (here we take $S=3$), the nonzero elements $c^i_{hk}$ are
	$$
	c^i_{i+1,k} =  \frac{S\,p_{i+1,k}}{r_{i+1} - r_i},
	\quad
	c^i_{i,k} = S\,\left( \dfrac{ p_{k,i}}{r_{i+1} - r_i} + \dfrac{p_{i,k}}{r_i - r_{i-1}}\right),
	\quad
	c^i_{i-1,k} = \frac{S\, p_{k,i-1}}{r_i - r_{i-1}}.
	$$	
	Finally, the full coefficients $B^i_{hk}$ are constructed 
	{combining the baseline terms and the correction terms as}
	$$
	B^i_{hk} = a^i_{hk} + c^i_{hk}.
	$$	
	}
	\\
	Beyond technical details, the analytical expression of this probability reflects the fact that the economic interactions among individuals determine the transfer of people from one functional subsystem to another. 
	We point out that more {elaborate} choices for both interaction frequencies and transition probabilities can be made to model other phenomena, such as socio-economic competition \cite{bellomo2008modeling, marsan2008towards}. 
	Nevertheless, we keep the model presented above in order to focus on the nonconservative terms. 
	
	In this closed-market society, the time-dependent proliferative/destructive rates $\mu_{hk}(t)$, for $h, k \in \{1, 2, \dots, n\}$, 
	stand for the birth/death rate due to the stochastic binary encounters between individuals of the $h$-th functional subsystem and those of the $k$-th functional subsystem. 
	Specifically, for our purposes, we consider only nonconservative rates related to occurring between pairs of individuals within the same subsystem.
	This means that 
	$$
	\mu_{hk}(t)= 0, \quad \forall h\neq k,\, \forall t>0.
	$$
	From an application viewpoint, we extend the meaning of the coefficient $\mu_{ii}(t)$, for $i \in \{1, 2, \dots, n\}$ and $t>0$, allowing it to represent birth/death natural events related to the $i$-th wealth-class. In other words, we suppose that the interaction among individuals of the same class may have both positive (reproduction) or negative (suppression) outcomes and that this dynamic has a cyclical behavior. Considering all of the above, and inspired by \cite{kendall1948generalized}, these coefficients are modeled through some suitable periodic functions. Hereafter, we assume
	\begin{equation}\label{muii}
		\mu_{ii}(t)=\bar\mu_{ii}\,\cos(2t),
	\end{equation}
	where coefficients $\bar{\mu}_{ii}$, for $i \in \{1,2, \dots, n\}$, are positive constants that model natural birth/death processes of the $i$-th wealth-class. 
	Furthermore, the function $\cos(2t)$ is the same for all functional subsystems. The coefficient $\bar\mu_{ii}$ is a weight of this process for people of the $i$-th income class, due to the specific wealth state. In particular, the coefficients $\bar\mu_{ii}$ are assigned as follows. For $n=3$,
	\begin{equation}\label{ParsMu1}
		\bar\mu_{11}=0.25,\quad \bar\mu_{22}=0.11,\quad \bar\mu_{33}=0.06 ,
	\end{equation}
	whereas, for $n=5$,
	\begin{equation}\label{ParsMu2}
		\bar\mu_{11}=0.25,\quad \bar\mu_{22}=0.16,\quad \bar\mu_{33}=0.11,\quad \bar\mu_{44}=0.08,\quad \bar\mu_{55}=0.05.
	\end{equation}
	In both cases $n=3$ and $n=5$, time-dependent proliferative/destructive rates decrease as income classes increase; there are two reasons for this. On the one hand, individuals in lower classes typically exhibit a higher tendency to have more children, resulting in a more pronounced population growth. 
	On the other hand, reciprocal support and caregiving among individuals of the same class may be lacking in lower economic strata.
	This tendency can be associated with the transmission of unhealthy lifestyle habits, the spread of diseases, and even real competition for survival.

	Finally, we introduce specific expressions for the coefficients $\lambda_i(t)$, for $i \in \{1, 2, \dots, n\}$, modeling the nonconservative external force field \eqref{forcshape}. More precisely, we want to use these terms to model effects and consequences of some slow or sudden events on a population. Of course, these consequences cannot be the same for each income class, as expected by experience and depicted by facts. Hereafter, we call these events ``shocks", where each 
	shock is characterized by some specific properties in terms of impact and duration. 
	It is worth underlining that this action is not directly related to interactions among individuals, but it is an external action on each income class. 
	
	As previously mentioned, the nonconservative term represents the impact of the shock on the society under consideration, which is viewed as part of a larger population. The shock induces the movement of individuals out of society but also accounts for their potential return. This can occur for different reasons, for example, the material destruction of houses and buildings after an earthquake, or the tendency to escape from an epidemic outbreak. Moreover, also the behavior of individuals some months or years after the shock may still be influenced by the shock itself. They may tend to come back or to still go away depending e.g. on the job availability or the reconstruction policies.
	
	\medskip
	
	Two different scenarios are considered in what follows, concerning the shape of the external force field $\mathbf{F}[\mathbf{f}](t)$. 
	
	\bigskip
	
	\noindent
	{\textit{Scenario 1 -- slow shock.}} 
	In the first scenario, we consider an external force field modeling a ``slow shock", i.e. an event that begins, grows, reaches its peak, and declines until its effect is no longer felt on the overall population.  For instance, a slow shock can be used for situations related to climate changes that are not sudden events, or for epidemic outbreaks that take a certain amount of time to reach their maximum expansion. Thus, the coefficients $\lambda_i(t)$, for $i \in \{1, 2, \dots, n\}$, are assumed to be
	\begin{equation}\label{shock1}
		\lambda_{i}(t)=\alpha_{i}\,\frac{1}{\sigma\sqrt{2\pi}}\exp\left(-\frac{(t-\nu)^2}{2\sigma^2}\right),
	\end{equation}
	where $\alpha_i \in \mathbb{R}$, for $i \in \{1, 2, \dots, n\}$. The slow shock \eqref{shock1} is modeled by using a Gaussian function centered at $t = \nu$, with variance $\sigma >0$. 
	The parameters $\nu$ and $\sigma$ denote the moment in time
	when the external field has the most pronounced impact on the population and the duration of time over which the shock affects the population, 
	respectively.
	The coefficients $\alpha_{i}$ model the impact and related consequences of the slow shock on each income class,
	since it is reasonable to assume that an external phenomenon acts differently and has a non-identical effect on each income class. 
	Then, for  $i \in \{1, 2, \dots, n\}$, a slow shock is modeled by the external force field
	$$F_i[\mathbf{f}](t)=\alpha_{i}\,\frac{1}{\sigma\sqrt{2\pi}}\exp\left(-\frac{(t-\nu)^2}{2\sigma^2}\right)f_i(t).$$
	Therefore, the dependence of the external action on the specific functional subsystem is expressed in \eqref{shock1} through coefficients $\alpha_i$, for $i \in \{1, 2, \dots, n\}$, whereas on the current state of the functional subsystem itself through the distribution function $f_i(t)$, for $i \in \{1, 2, \dots, n\}$.
	
	\bigskip
	
	\noindent
	{\textit{Scenario 2 -- sudden shock.}} 
	In a second scenario, we consider an external force field modeling a ``sudden shock", i.e. an event that 
	instantaneously appears as a sudden peak with no smooth initial phase, and then decreases until its effect is no longer felt.  For instance, a catastrophic event such as an earthquake or a hurricane may be modeled in this way.   Hence, the coefficients $\lambda_i(t)$, for $i \in \{1, 2, \dots, n\}$, are assumed to be
	\begin{equation}
		\label{shock2}
		\lambda_{i}(t)=\alpha_{i}\,\frac{1}{\sigma\sqrt{2\pi}}\exp\left(-\frac{(t-\nu)^2}{2\sigma^2}\right)\mathbf{I}_{\{t\geq\nu\}},
	\end{equation}
	with $\alpha_i \in \mathbb R$, for all $i \in \{1, 2, \dots, n\}$. The sudden shock \eqref{shock2} is modeled through a Gaussian function centered at $t=\nu$, with variance $\sigma >0$, as in the previous case \eqref{shock1} of slow shock, but now, the sudden behavior is present in \eqref{shock2} trough the indicator function $\mathbf{I}_{\{t\geq\nu\}}$ that mimics the instantaneous and abrupt change on the behavior. Then, for all $i \in \{1, 2, \dots, n\}$, a sudden shock is modeled by the external force field
	$$
	F_i[\mathbf{f}](t)=\alpha_{i}\,\frac{1}{\sigma\sqrt{2\pi}}\exp\left(-\frac{(t-\nu)^2}{2\sigma^2}\right)\mathbf{I}_{\{t\geq\nu\}}f_i(t).
	$$
	For the purposes of our model, the proliferative/destructive rates $\mu_{hk}$, for $h,k \in \{1, 2,\dots, n\}$, will be fixed for both configurations as in \eqref{ParsMu1} and \eqref{ParsMu2}. 
	The focus of our investigation will be, then,  on the way how the specific shape of the external force field (slow or sudden shock) influences 
	the dynamics of the overall economic system.

	It is worth noting that the above choice for the external force field in terms of a Gaussian function represents just the first possibility for modeling shock events that impact a socio-economic system. Specifically, the selection of the Gaussian shape is primarily driven by its distinctive characteristics, such as the occurrence of peaks at a specific time, the intensity, and the time duration.
	
	Some analytical considerations are here required regarding the selection of nonconservative parameter shapes, 
	i.e. proliferative/destructive rates $\eta_{hk}(t)$, for $h,k \in \{1, 2,\dots, n\}$, and the external force field $\mathbf{F}[\mathbf{f}](t)$. 
	If a slow shock \eqref{shock1} occurs, the assumptions of Theorem \ref{thm1} are satisfied; then, given an initial data $\mathbf{f}^0\in\left(\mathbb{R}^+\right)^n$, there exists a unique and positive solution, locally in time. 
	Indeed, the slow shock \eqref{shock1} is modeled by continuous functions, regular enough for Theorem \ref{thm1}. 
	Conversely, if a sudden shock \eqref{shock2} occurs, the situation is slightly different. In this latter case, in fact, the coefficients $\lambda_{i}(t)$ are modeled by noncontinuous functions, due to the presence of the indicator function $\mathbf{I}_{\{t\geq\nu\}}$. 
	Therefore, the results of Theorem \ref{thm1} do not apply in this case. Anyway, due to the choice of coefficients $\lambda_{i}(t)$, for $i \in \{1, 2, \dots, n\}$, and the expression of proliferative/destructive rates \eqref{ParsMu1} and \eqref{ParsMu2} we can, at least, conclude that there exists a solution, locally in time, by using some standard results of theory of ordinary differential equations, such as Carath\'eodory's theorem (see \cite{coddington2012introduction} for details). Indeed, all coefficients of equation \eqref{eqkin} are bounded by measurable functions for each fixed time $t>0$. 
	It is worth emphasizing that in the case of a sudden shock, the uniqueness of the solution is not guaranteed in general. Nevertheless, the solution remains positive due to the structure of the kinetic system \eqref{eqkin} itself. 
	
	{Then, the kinetic model \eqref{eqkin2} rewrites, for $i \in \{1, 2, \dots, n\}$,
		\begin{equation}\label{eqkinn}\begin{split}
				\frac{df_i}{dt}(t)&=\sum_{h,k=1}^{n}B^i_{hk}f_h(t)f_k(t)-f_i(t)\sum_{k=1}^{n}f_k(t)\\&
				+ f_i^2(t)\,\bar\mu_{ii}\cos(2t) +\alpha_{i}\,\frac{1}{\sigma\sqrt{2\pi}}\exp\left(-\frac{(t-\nu)^2}{2\sigma^2}\right)\mathbf{I}_{\{t\geq\nu\}}f_i(t) .
			\end{split}
	\end{equation}
	According to Theorem \ref{thm1}, system \eqref{eqkinn} can be split, for $i \in \{1, 2, \dots, n\}$, as
	
	\begin{equation}\label{eqth21}
		\frac{df_i}{dt}(t)=J_i[\mathbf{f}](t)+f_i(t)\,S_i[\mathbf{f}](t),
	\end{equation}
	where
	\begin{align*}
		J_i[\mathbf{f}](t)&=G_i[\mathbf{f}](t)-L_i[\mathbf{f}](t)\\
		S_i[\mathbf{f}](t)&=\left(\bar\mu_{ii}\cos(2t)f_i(t)+\alpha_{i}\,\frac{1}{\sigma\sqrt{2\pi}}\exp\left(-\frac{(t-\nu)^2}{2\sigma^2}\right)\mathbf{I}_{\{t\geq\nu\}}\right).
	\end{align*}
	This structure ensures the preservation of the nonnegativity of the solution, as proved in Theorem \ref{thm1}. 
	The operator on the right-hand side of the \eqref{eqth21} is a bounded Lipschitz operator, even if with a discontinuity. 
	Carath\'eodory's theorem \cite{coddington2012introduction} 	
	{then guarantees the existence of} at least a positive local solution.}

	{\begin{remark}\label{rem1}
			The particular choice of the birth/death coefficients as in \eqref{muii}, together with the coefficients for the shock given in \eqref{shock1} or \eqref{shock2} and the interaction rates in \eqref{parmetri conservativi}, may provide conditions under which global solutions are expected to exist, thereby extending the analytical insight provided in Theorem \ref{thm1}. However, the derivation of a rigorous global result in a more general framework goes beyond the scopes of this work. {Ne}vertheless, we present below some considerations that could lead to such a result, at least in the particular situation here studied.
			
			If one sums all the equations in the final system \eqref{eqth21} over $i \in \{1, 2, \dots, n\}$, it follows that
			\begin{equation}
				\begin{aligned}
					\frac{d}{dt}\sum_{i=1}^n f_i(t) &= \! \sum_{i=1}^n f_i(t)\, S_i[\mathbf{f}](t)
					\label{eqth22}
					\\[2mm]
					&
					= \! \left(\sum_{i=1}^n\bar\mu_{ii}f_i^2(t) \! \right) \! \cos(2t) \!+\! 
					\left(\sum_{i=1}^n\alpha_i f_i(t)\right) \! \exp\left( \! -\frac{(t-\nu)^2}{2\sigma^2}\right)\mathbf{I}_{\{t\geq\nu\}}.
				\end{aligned}
			\end{equation}
			Now, let us focus on the right-hand side of the equation \eqref{eqth22}. The second term rapidly decreases due to the presence of the exponential function. Regarding the first term, since the functions $f_i$'s are nonnegative, along with the coefficients $\bar{\mu}_{ii}$, we have
			$$\sum_{i=1}^n \bar{\mu}_{ii}f_i^2(t) > 0.$$
			Therefore, the coefficient that involves the cosine changes sign during the time evolution. Accordingly, the derivative of the total population $\displaystyle \sum_{i=1}^{n} f_i(t)$ appears to have an alternating sign, suggesting that the total population cannot grow indefinitely during the evolution of the system. Indeed, assuming a positive initial data $\mathbf{f}^0$, i.e. $ f_i^0 \geq 0 $ for all $ i \in \{1, 2, \dots, n\} $, the Cauchy problem related to system \eqref{eqkinn} may admit a global nonnegative solution $\mathbf{f}(t)$, i.e. for all $ t > 0 $. This observation may contribute to ensuring the global well-posedness of the associated Cauchy problem for system \eqref{eqkinn}, as further supported by the numerical simulations presented in the following sections.
	\end{remark}}

	%%%%%%%%%%%%%%%%
	
	\subsection{Numerical Simulations}
	\label{ssecnum}
	
	This subsection aims to illustrate the behavior of the model previously proposed by means of numerical simulations, allowing us to better appreciate the analytical properties previously discussed. For this purpose, the parameters will be chosen arbitrarily. We present results for scenarios 1 and 2, considering both cases of $n=3$ and of $n=5$ income classes.

	\bigskip
	
	\noindent
	\textit{Initial data.}
	The initial data $\mathbf{f}^0$ is definitively fixed for both cases. 
	In particular, with $n=3$ wealth-classes, it is set
	\begin{equation}\label{InData1}
		\mathbf f^0=(0.39,0.46,0.15),
	\end{equation}
	whereas, for $n=5$,
	\begin{equation}\label{InData2}
		\mathbf f^0=(0.08,0.47,0.32,0.1,0.03).
	\end{equation}
	It is worth stressing that this specific choice for the initial data reflects the idea of a nonsymmetric structure of income classes in a closed-market society here considered. Indeed, for the case of $n=3$, the total amount of lower and middle-class individuals represents the majority of the population. Analogously, with $n=5$, the lower, lower-middle, and middle classes together stand for the majority, if compared to the upper-middle and upper classes. The reason for this choice for the initial data is twofold. On the one hand, we aim at depicting a society where the lower and middle classes predominate; this appears more clearly in the case with $3$ income classes. On the other hand, we expect few ``very rich" people, and this is outstanding in the case with $5$ income classes, where $f^0_5=0.03 \ll f^0_i$, for $i \neq 5$. { Finally, we underline that in both cases, the initial data is such that
		$$\sum_{i=1}^{n}f^0_i=1,\qquad n\in\{3,\,5\}.$$
		This last choice is related to the fact that, if we consider the conservative framework, i.e. the system \eqref{eqkin} without time-dependent proliferative/destructive rate $\mu_{hk}(t)$, for $h,k \in \{1, 2, \dots, n\}$, and external force field $\mathbf{F}[\mathbf{f}](t)$, it is possible to find for the Cauchy problem a unique and positive solution $\mathbf{f}(t)$, globally in time, in such a way that
		$$\sum_{i=1}^{n}f_i(t)=1, \qquad \forall t>0.$$
		Thus, the solution can be interpreted as a probability.}
	
	\bigskip
	
	\noindent
	\textit{External force field.} 
	The parameters $\nu$ and $\sigma$ that characterize the Gaussian structure of the external force field
	for both scenarios 1 and 2 are set as $\nu=50$ and $\sigma=20$.
	Given the form of the coefficients $\lambda_i(t)$ for a slow shock \eqref{shock1},
	this means that the peak of the event is gained at $t=50$.
	Similarly, for a sudden shock \eqref{shock2}, the event starts in its peak at $t=50$.
	
	Concerning the coefficients $\alpha_i$, for $i \in \{1, 2, \dots, n\}$, 
	we assign, for the case $n=3$, the values 
	\begin{equation}
		\label{ParsAl1}  
		\alpha_1=-0.25,\quad \alpha_2=-0.15,\quad \alpha_3=-0.05.
	\end{equation}
	Bearing the structure of the external force field in mind, this choice provides a shock,
	for both slow and sudden cases, that has a greater impact on the lower classes, with respect to the upper ones. 
	Analogously, in the case $n=5$, the values decrease from the lower class to the upper one. 
	Specifically:
	\begin{equation}
		\label{ParsAl2}  
		\alpha_1=-0.25,\quad \alpha_2=-0.2,\quad \alpha_3=-0.15,\quad \alpha_4=-0.1,\quad \alpha_5=-0.05.
	\end{equation}
	This choice for the values of coefficients $\alpha_i$, in \eqref{ParsAl1} and \eqref{ParsAl2}, suitably represents the role played by the external force field $\mathbf{F}[\mathbf{f}](t)$ in the present work; according to what expected, we suppose that the impact is bigger for lower classes compared to upper ones.
	
	For example, natural catastrophes tend to have a more significant impact on lower-income classes, as they often reside in overcrowded areas with inadequate sanitary conditions. Additionally, public rescue services do not always manage to meet the needs of the entire population in these areas.  In addition,  we suppose that the shock influences each income class negatively, i.e. $\alpha_{i}\leq 0\,\forall\, i=1,\ldots,n$. Despite this assumption being analytically convenient, it may not fully align with real-world situations,  as we will see afterward.
	
	\bigskip
	
	In our simulations, we consider both scenario 1 of slow shock and scenario 2 of sudden shock introduced in Subsection \ref{ssecks}.
	Then, we build two further complex scenarios of consecutive shocks, the first one being a sudden shock and the second one
	being either a slow or a sudden shock.
	To better apprehend the impact of the shocks in the evolution of populations, we
	start with the numerical investigation of a baseline scenario in which no external force field acts on the overall system, i.e. the populations are not exposed to any shock.
	
	In all scenarios, we are interested in the time-evolution of the distribution functions $f_i(t)$, 
	for both cases of $n=3$ and $n=5$ income classes. 
	
	The main objective is to analyze the effects of the shocks on the classes and how they react to these external force fields. Some relevant questions are: Do the classes restore after a shock?
	Do they return to their previous configuration?
	
	We consider the reference time window $[0,\,200]$ and perform the simulations using standard MATLAB routines 
	for solving ordinary differential equations, 
	i.e. fourth-order Runge-Kutta methods.
	
	%%%%%%%%%%%%%%%%
	
	\subsubsection{Baseline scenario}
	\label{ssecBS}
	
\begin{figure}[ht!]
		\centering
		\includegraphics[scale=0.143]{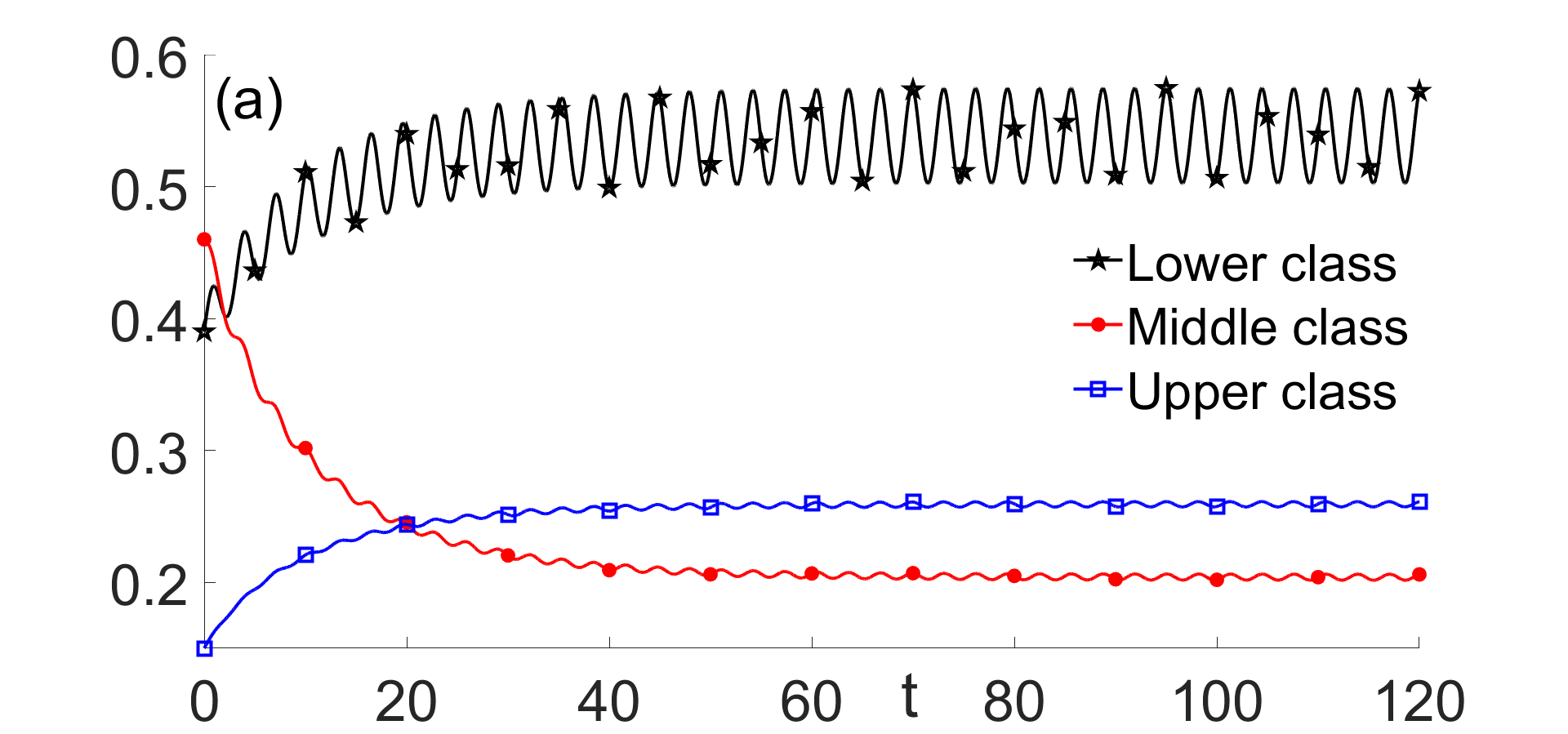}
		\includegraphics[scale=0.143]{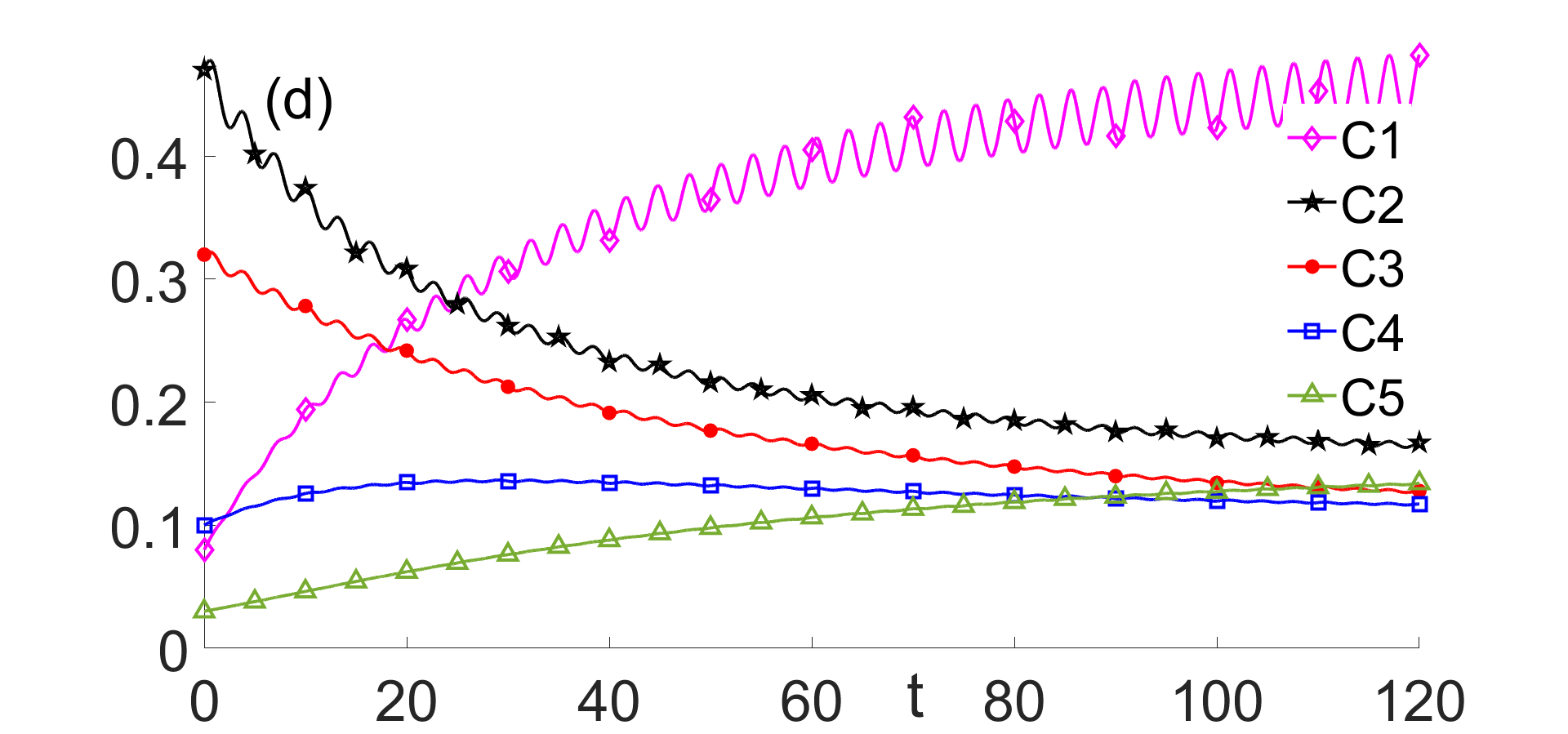}\\
		\includegraphics[scale=0.143]{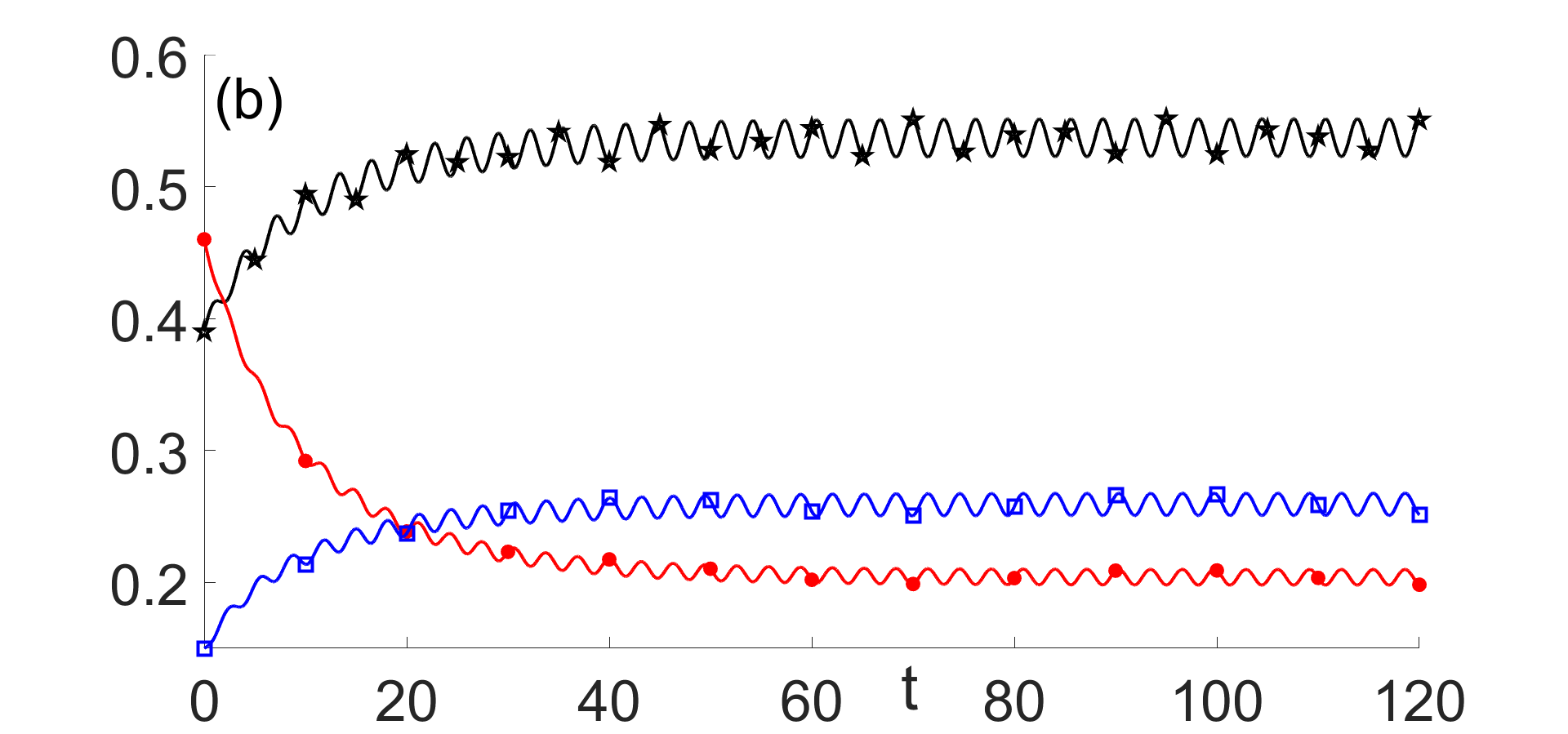}
		\includegraphics[scale=0.143]{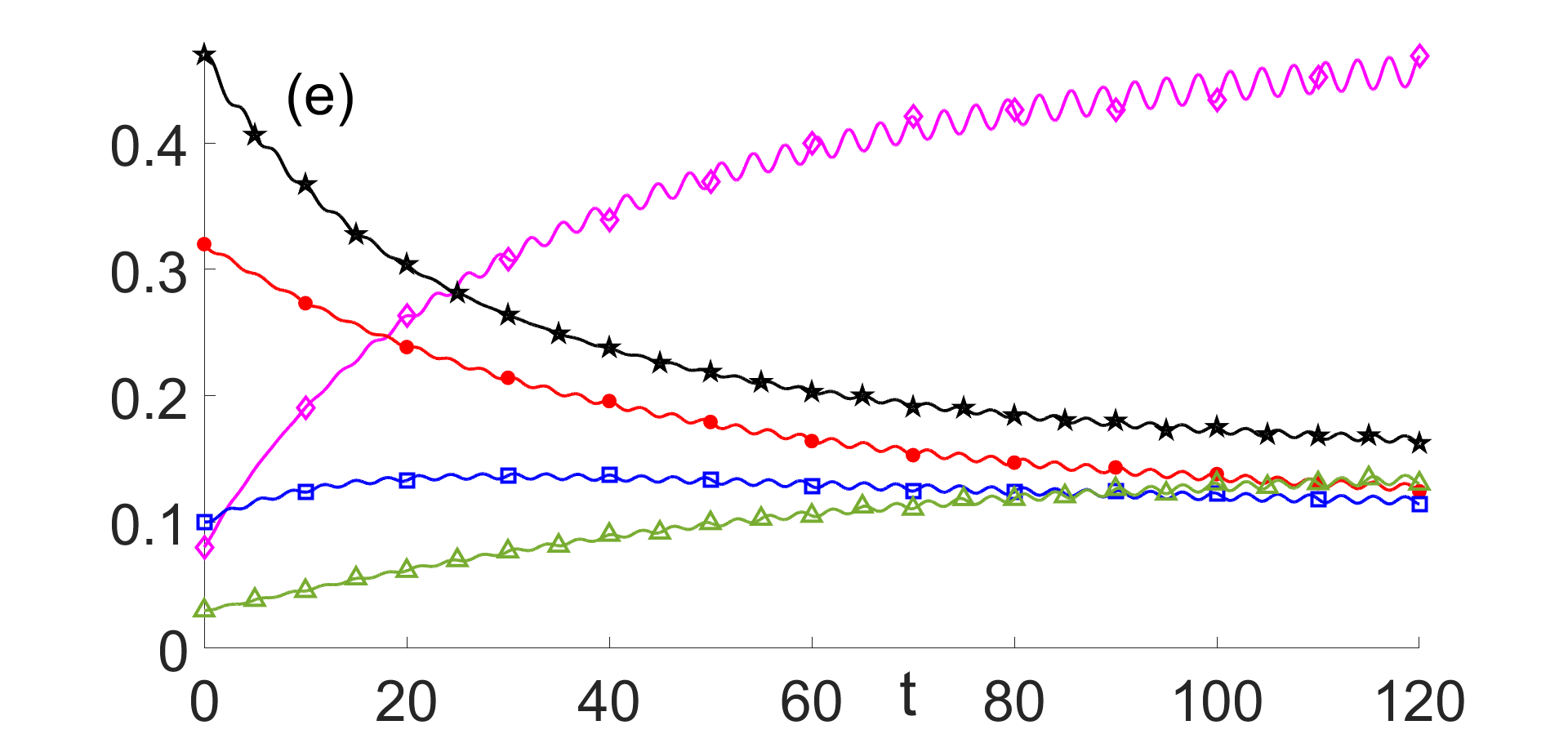}\\
		\includegraphics[scale=0.143]{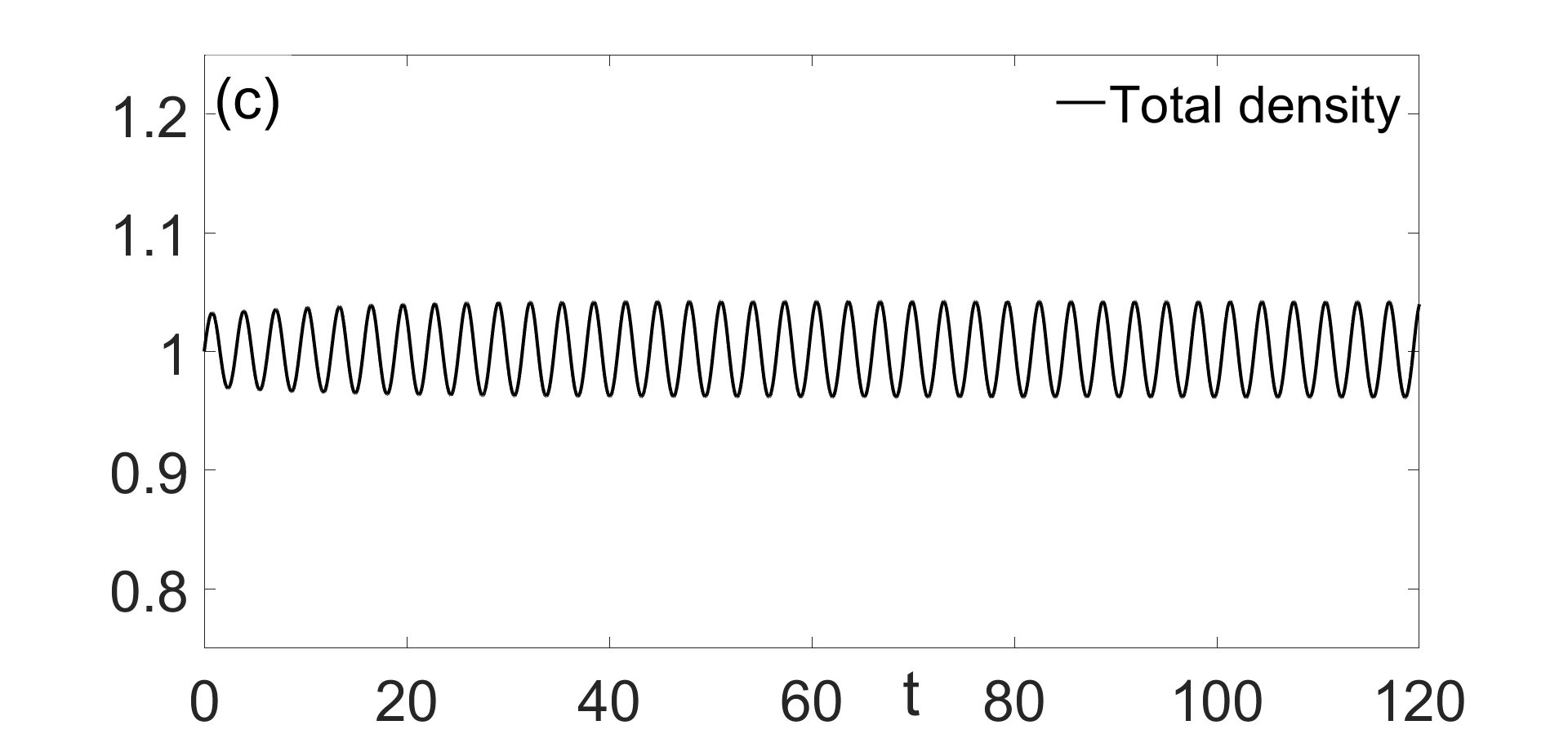}
		\includegraphics[scale=0.143]{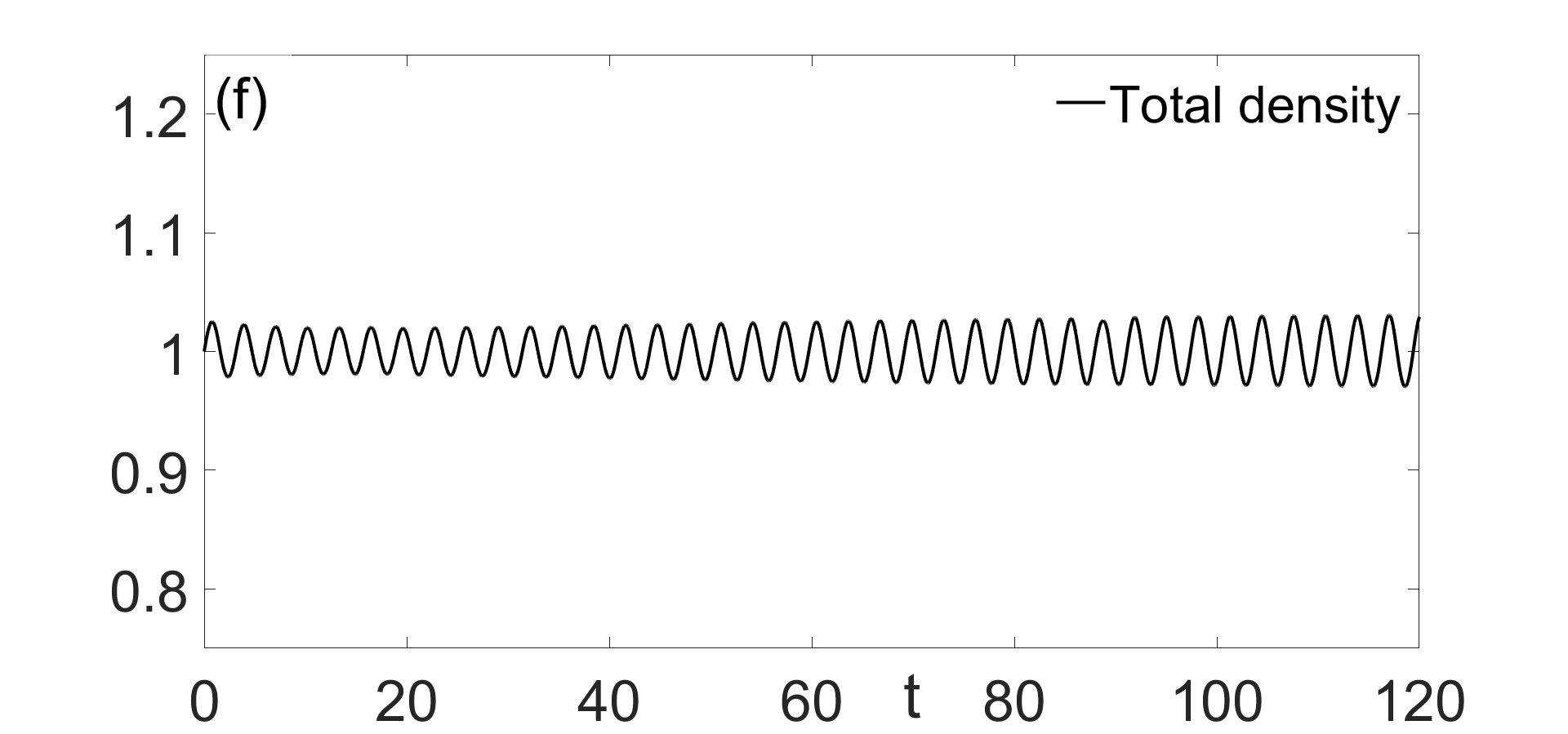}\\
		\caption{{{\it Baseline scenario.} Time evolution of the population distribution functions without undergoing any shock. 
				{Left column: three-classes population with parameters 
				as in \eqref{parmetri conservativi}-\eqref{ParsMu1}-\eqref{ParsAl1}
				and initial data \eqref{InData1}. 
				Right column: five-classes population with parameters 
				as in \eqref{parmetri conservativi}-\eqref{ParsMu2}-\eqref{ParsAl2} and initial data \eqref{InData2}. 
				Panels {(a) - (d)}: distribution functions; 
				panels {(b) - (e)}: ratio between distribution functions and total density; 
				panels {(c) - (f)}: total density.}}}
		\label{NoEvent}
	\end{figure}

	In this scenario, we solve the kinetic system \eqref{eqkin} with $\mathbf{F}[\mathbf{f}](t)=0$, for all $t \geq 0$. 
	Therefore, only conservative and nonconservative binary interactions occur. 
	The behavior of the distribution functions $f_i(t)$ is displayed in Figure \ref{NoEvent}.
	{We} {report, in the left column, the case with $n=3$, while in the right one the case with $n=5$.
	In the first row the total amount of each income class for the two cases is displayed, 
	while in the second row we show the ratio between each  income class and the total density, i.e.
	$f_i(t)/\rho(t),\,i=1,\ldots,n$; the total density $\rho(t)$, in turn, is shown in the last row.}
	In panels {(a), (b)} we observe how the lowest class (black-starred line) undergoes a higher oscillation and tends to increase rapidly. 
	Also, an increase in the upper class (blue-squared line) is appreciable, while the middle class (red-dotted line) is decreasing. 
	These simulations show that, in a closed-market society, the number of individuals of the middle class tends to decrease, the number of lower-class individuals increases, and there is a sort of polarization between lower classes and upper classes. 
	Indeed, we are not taking into account any redistribution mechanism through taxes, \cite{bertotti2010modelling}. 
	{From a modeling point of view, this is due to the choice of transition probabilities $B_{hk}^i$. In fact, {let} us define, for each income class $i$, the quantities $$\Phi_i={\displaystyle\sum_{k=1}^n B_{ik}^i},\quad\Psi_i={\displaystyle\sum_{\substack{
					h,k=1 \\
					h \neq i
			}}^n B_{hk}^i}\bigg/{\displaystyle\sum_{\substack{
					h,k=1 \\
					h\neq i
			}}^n B_{ik}^h},
	$$
	where $\Phi_i$ accounts for the individuals who remain in the $i$-th class after interactions with individuals from other classes, and $\Psi_i$ represents the net balance between those entering the $i$-th class and those leaving it. For the present case, we have the amounts reported in the table below. 
	\begin{table}[ht!]
		\centering{
		\begin{tabular}{|c||c|c|c|}
			\hline
			&Lower&Middle &Upper\\
			\hline
			\hline
			$\Phi$ & $2.95$ & $2.76$ & $2.92$\\
			\hline
			$\Psi$ & $2.18$ & $0.56$ & $1.54$\\
			\hline
		\end{tabular}}
		\end{table}
		\\
These results indicate that individuals in the lower and upper income classes are more likely to remain in their respective classes after interactions, compared to those in the middle class, as shown by the higher values of $\Phi$. Furthermore, the relatively high values of $\Psi$ for the lower class suggest that it receives a larger inflow of individuals from other classes, making downward transitions more common than upward ones in this setting. 
We also {note} that this tendency is {observed even} without {accounting for} the proliferative/destructive rates $\mu_{hk}(t)$ \cite{menale2023kinetic, menale2024kinetic}. 
}
	It is worth noting that the stronger oscillations of these latter individuals are related to the specific choice of non-conservative rates \eqref{ParsMu1}, 
	reflecting the fact that in the lower classes, the natality is higher, but also the mortality. 
	This is due to several reasons, among others less possibility to benefit from the health care system. 
	Moreover, the upper class increases up to a value that stabilizes with very small oscillations. 
	We may conclude, then, that in such an economic system both the lower and the upper class increase, while the middle class is drastically reduced. {We also observe how the total density oscillates around the value $\overline\rho=1$.}
	
	In panels {(d) - (e) - (f)}, where the case with $n=5$ is presented, we can observe certain similarities with the previous case for $n=3$ populations. 
	In particular, the lower class appears to dominate in the long-term behavior, while the upper class shows an increasing trend relative to the other classes. 
	It is noticeable that the 2nd, 3rd, and 4th functional subsystems exhibit a tendency towards particularly lower levels, contrasted by a significant and rapid growth of the lower class (indicated by the magenta-with-diamonds line) and a growth, albeit at a slower pace, of the upper class (green-with-triangles line). 
	This behavior provides a more detailed representation of what was already observed in the case with $n=3$. 
	In fact, a significant portion of the population transitions into the lower class, while individuals in the upper class consolidate their positions. 
	{This tendency is confirmed by the values of $\Phi$ and $\Psi$ that in this case are given in the table below.
		\begin{table}[ht!]
			\centering{
			\begin{tabular}{|c||c|c|c|c|c|}
				\hline
				&C1 & C2 & C3 & C4 & C5\\
				\hline
				\hline
				$\Phi$ & $4.96$ & $4.79$ & $4.74$ & $4.75$ & $4.91$\\
				\hline
				$\Psi$ & $2.76$ & $0.8$ & $0.86$ & $0.87$ & $1.45$\\
				\hline			
			\end{tabular}}
		\end{table} }\\ 
	The oscillatory pattern, similar to what is {observed} in the case with $n=3$, is still appreciable and is related again to the nonconservative rates \eqref{ParsMu2}. {However,} {in this case, the oscillations appear to be more damped, 
	{suggesting} that the increased complexity of the model {offers a} better {representation of} reality.} 
	Again, the strongest oscillations concern the lower class, while they are considerably weaker for the other classes.

	%%%%%%%%%%%%%%%%
	
	\subsubsection{Slow shock scenario}
	\label{ssecS1}
	
\begin{figure}[ht!]
		\centering	
		\includegraphics[scale=0.143]{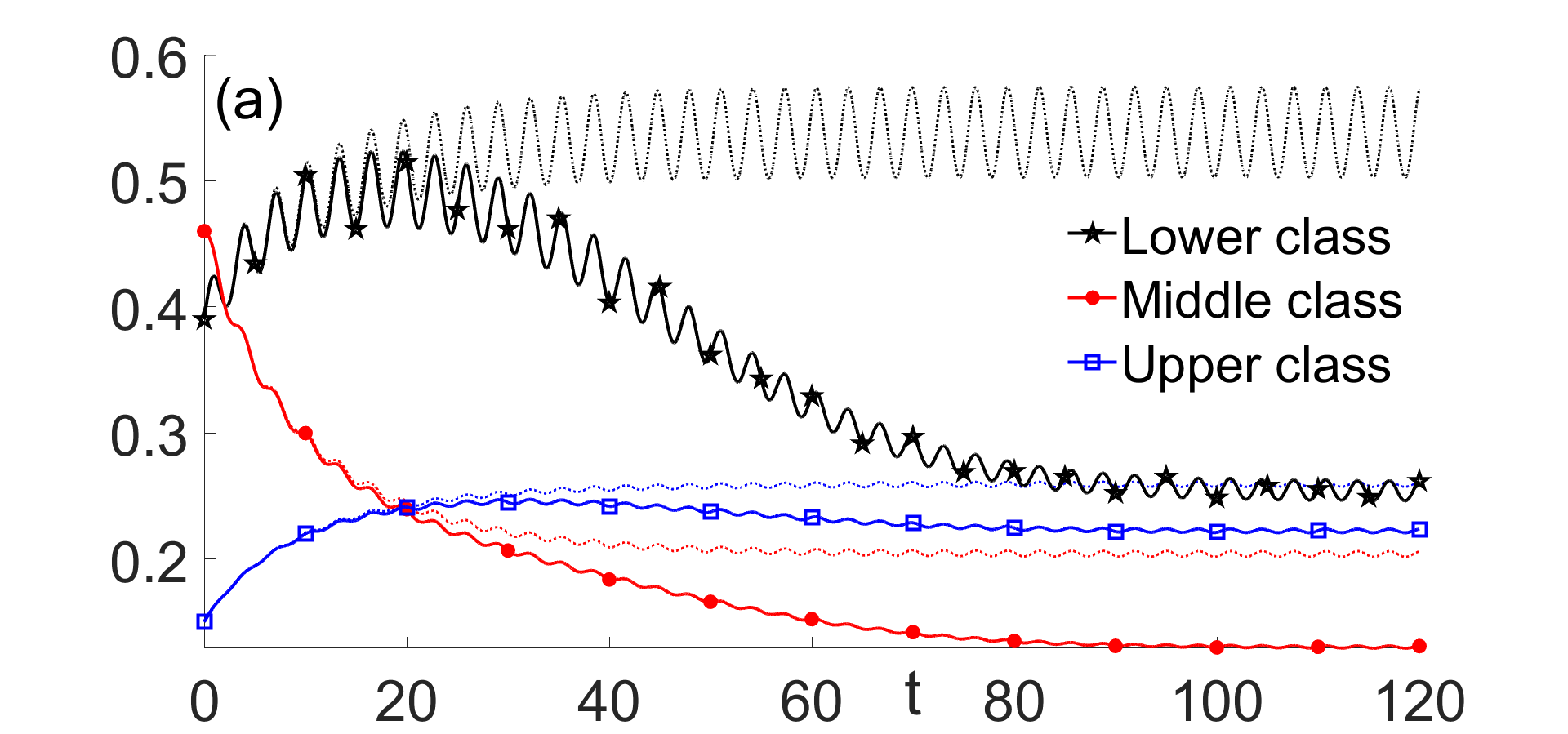}
		\includegraphics[scale=0.143]{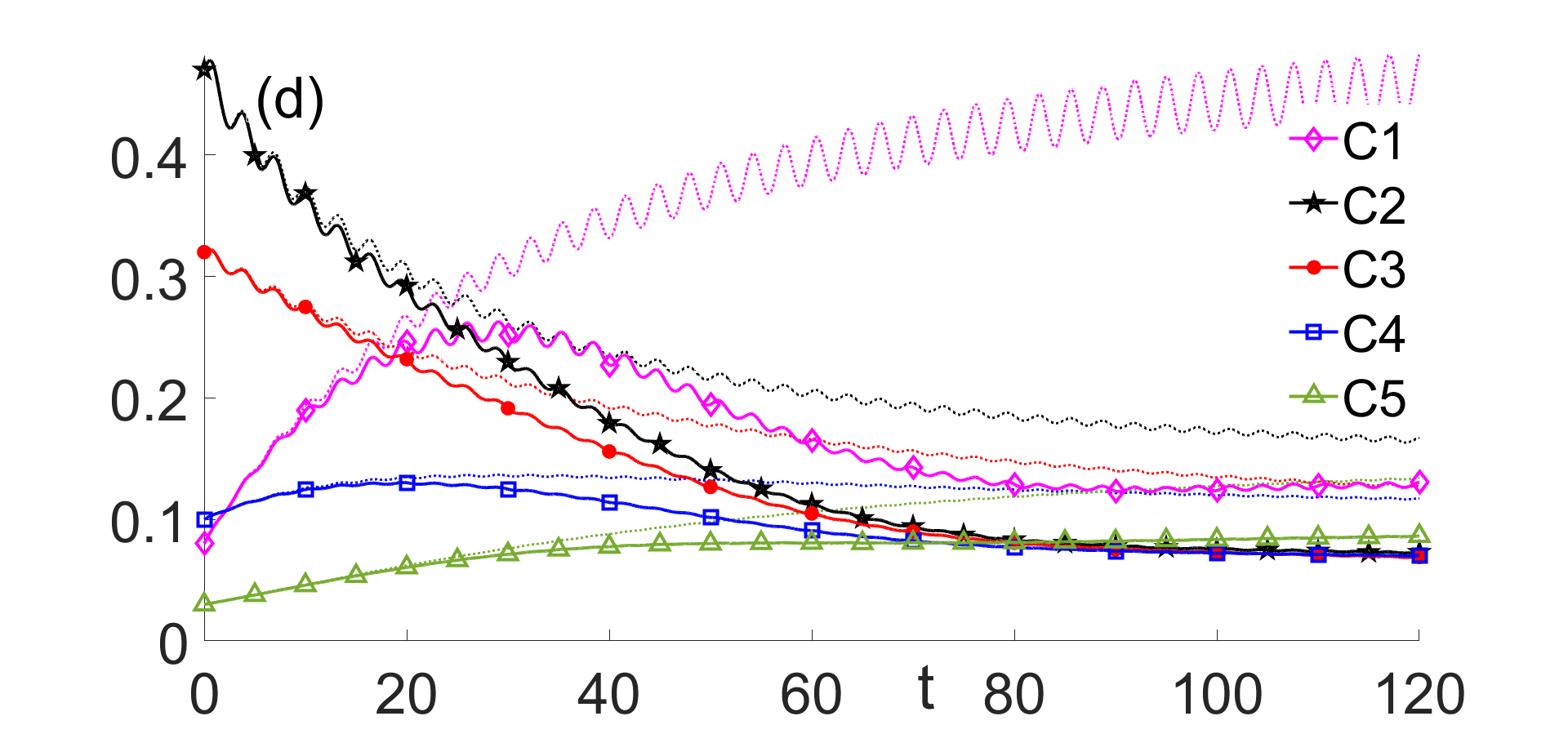}\\
		\includegraphics[scale=0.143]{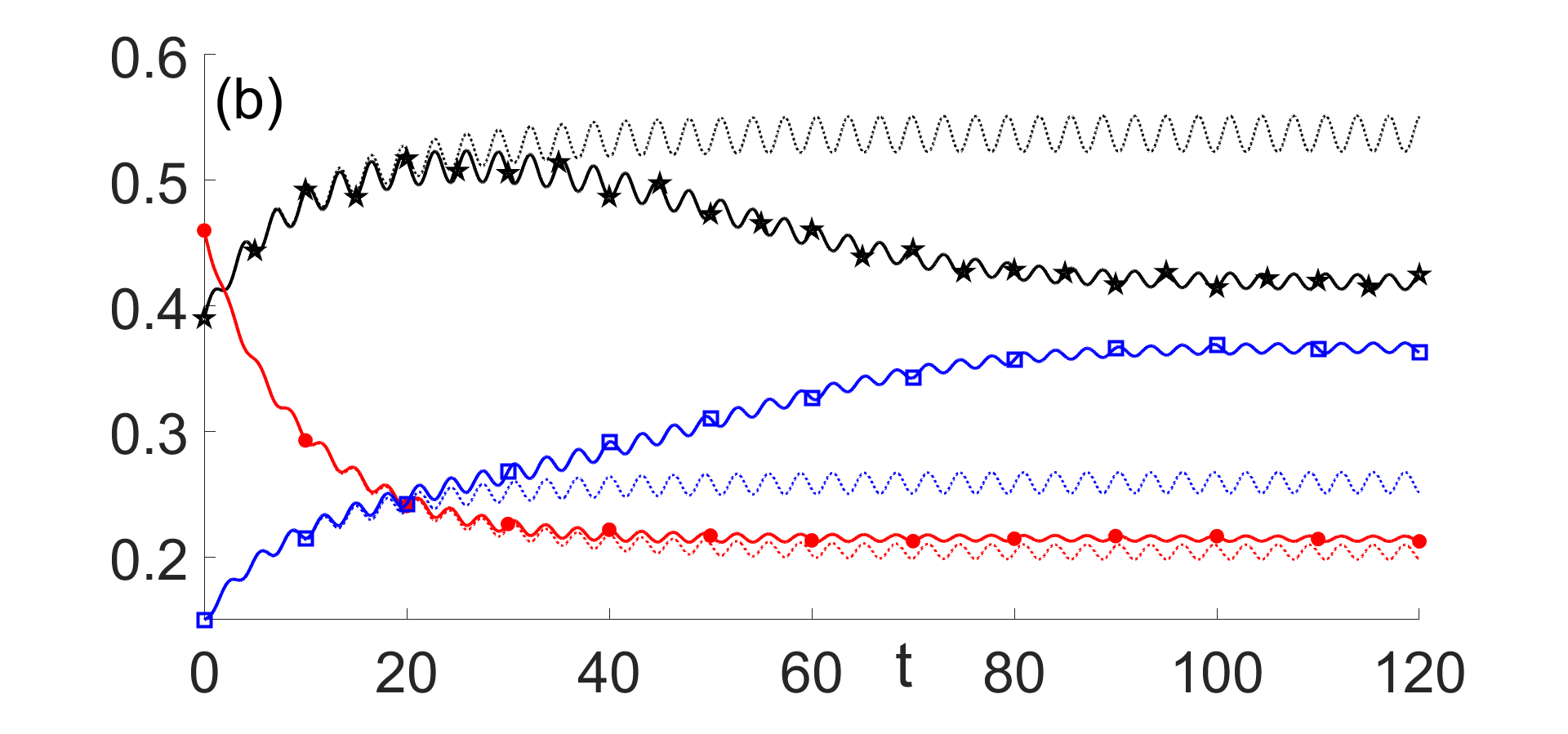}
		\includegraphics[scale=0.143]{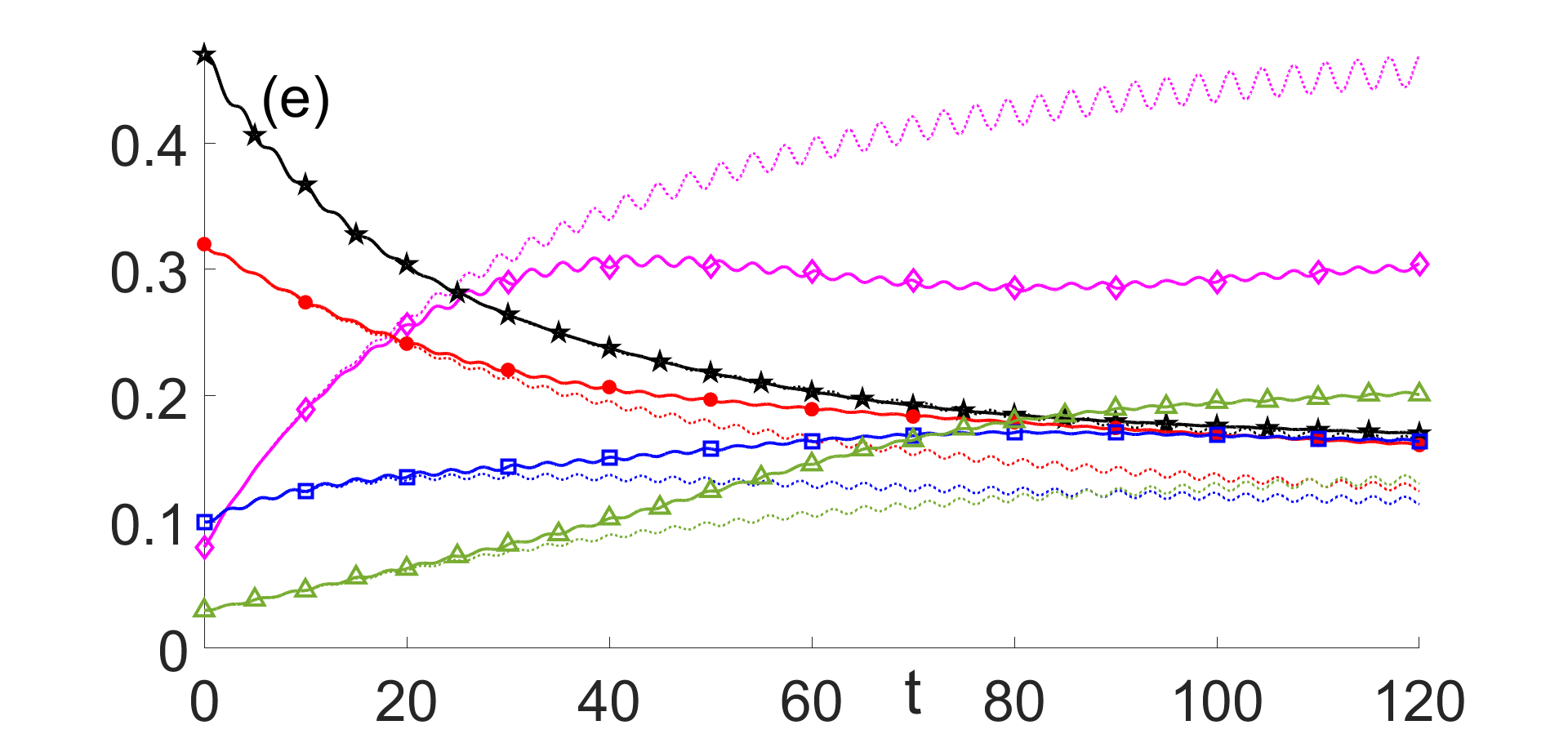}\\
		\includegraphics[scale=0.143]{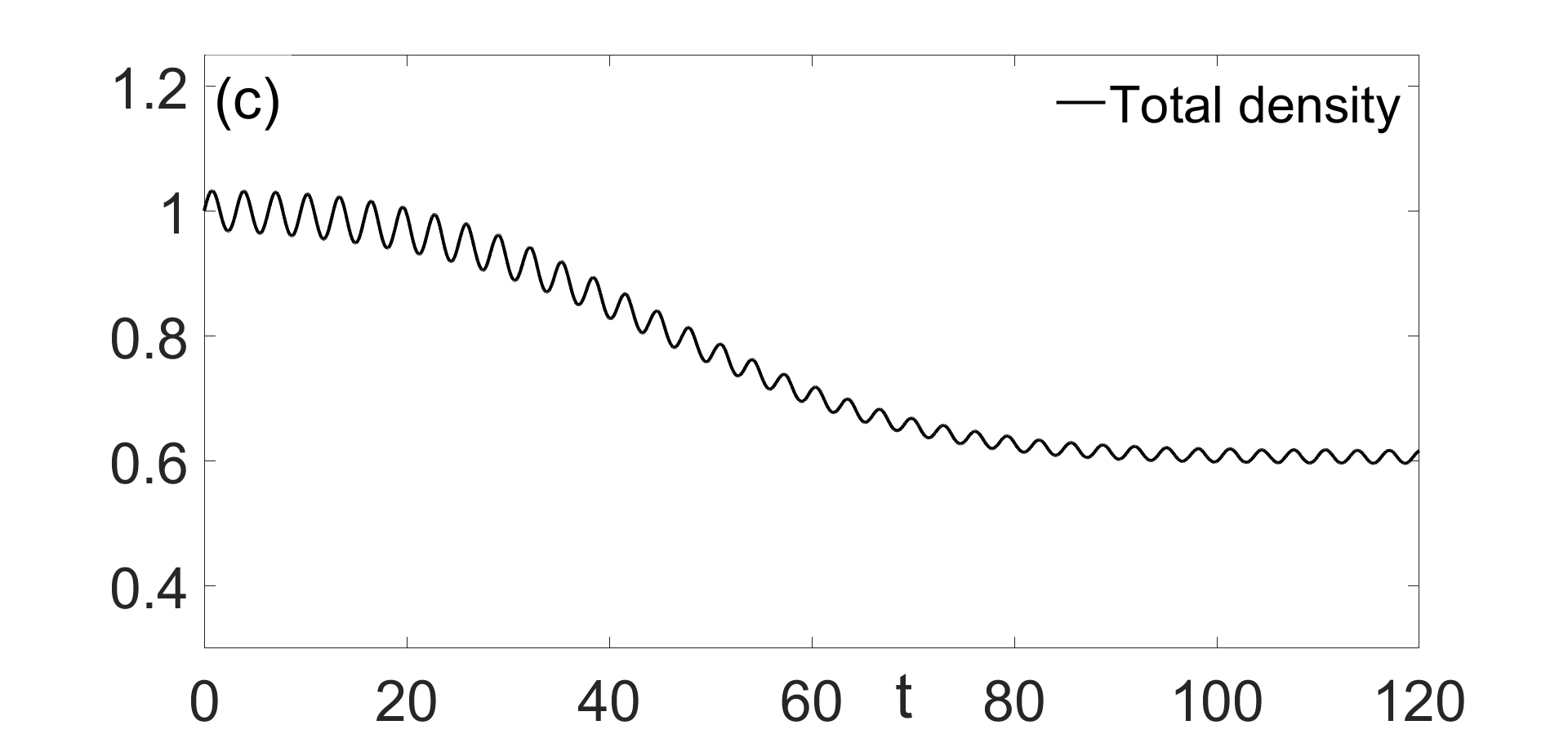}
		\includegraphics[scale=0.143]{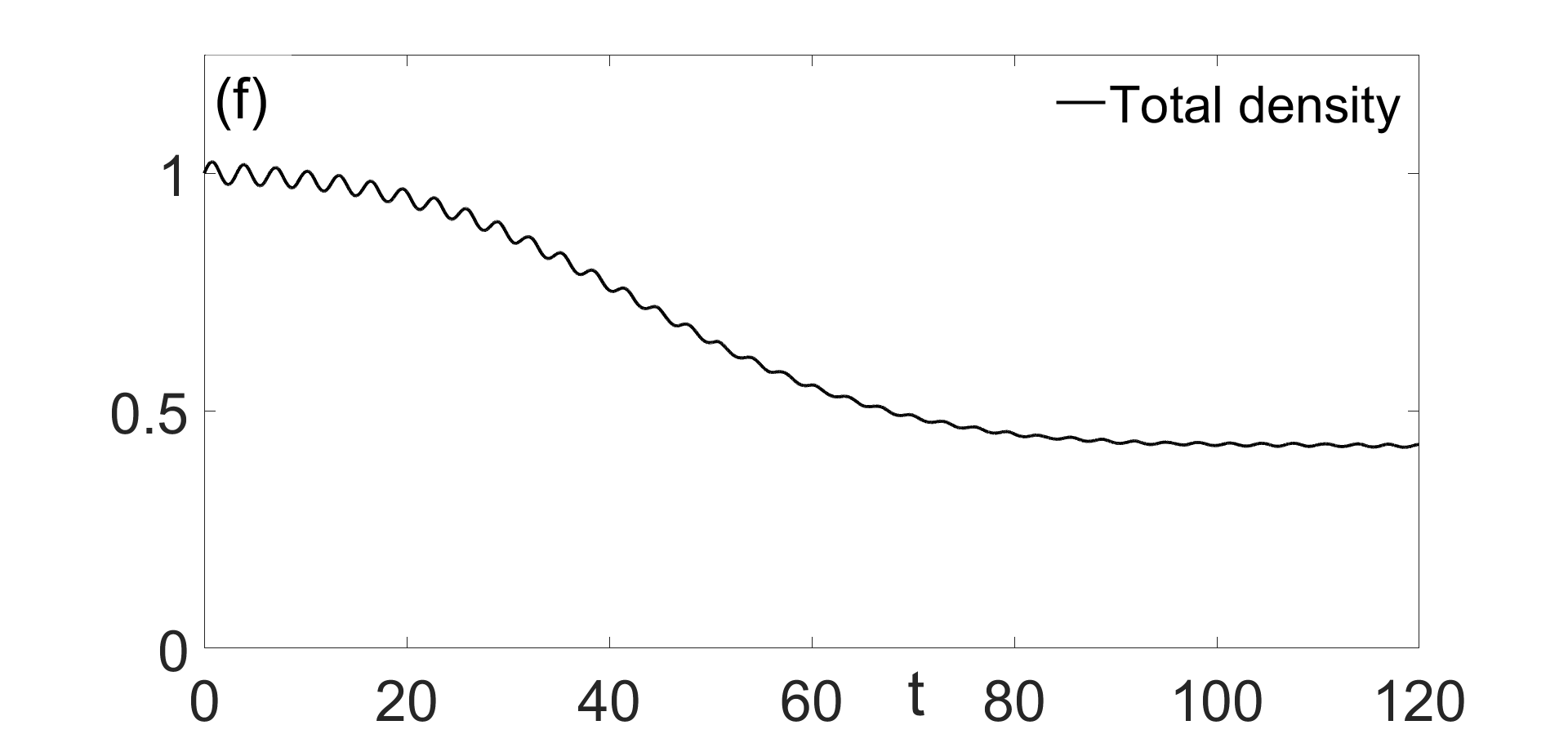}\\
		\caption{\textit{Slow shock scenario.}
			Time evolution of the population distribution functions when a slow shock occurs at $t=50$ with $\sigma=20$, compared to the reference case (dotted lines). {Left column: three-classes population with parameters {as in \eqref{parmetri conservativi}-\eqref{ParsMu1}-\eqref{ParsAl1}}
				and initial data \eqref{InData1}. 
				Right column: five-classes population with parameters as in \eqref{parmetri conservativi}-\eqref{ParsMu2}-\eqref{ParsAl2}
				and initial data \eqref{InData2}. 
				Panels {(a) - (d)}: distribution functions; 
				panels {(b) - (e)}: ratio between distribution functions and total density; 
				panels {(c) - (f)}: total density.}}
		\label{Crisis}
	\end{figure}

	As a first scenario depicting how an external force may impact society, we consider the one in which a slow shock, modeled by \eqref{shock1}, occurs. 
	The shock reaches its peak at $t=50$, and the coefficients $\alpha_i$ are chosen 
	according to \eqref{ParsAl1} and \eqref{ParsAl2} for the case $n=3$ and $n=5$, respectively. 
	The related simulations can be observed in Figure \ref{Crisis}.
	In particular, the case $n=3$ is shown in panels (a) - (b) - (c), and the case $n=5$ in panels (d) - (e) -(f). 
	The results are shown in comparison with the baseline scenario (represented by dotted lines). 
	Moreover, we report in Table \ref{tab1} the values of the densities for each class after the time $t=200$, 
	both for slow shock and baseline dynamics. 
		\begin{table}[ht!]
		\centering
		\begin{tabular}{|c||c|c|c|c|}
			\hline
			&Lower&Middle &Upper&Total\\
			\hline
			\hline
			Baseline&$0.507$&$0.202$&$0.257$&$0.966$\\
			\hline
			Slow shock&$0.246$&$0.129$&$0.221$&$0.596$\\
			\hline
			Ratio &$0.486$&$0.639$&$0.859$&$0.617$\\
			\hline
		\end{tabular}
		\vskip 0.3cm
		\begin{tabular}{|c||c|c|c|c|c|c|}
			\hline
			&C1 & C2 & C3 & C4 & C5&Total\\
			\hline
			\hline
			Baseline &$0.456$&$0.15$&$0.111$&$0.11$&$0.145$&$0.972$\\
			\hline
			Slow shock&$0.139$&$0.066$&$0.06$&$0.065$&$0.093$&$0.423$\\
			\hline
			Ratio&$0.305$&$0.437$&$0.536$&$0.592$&$0.643$&$0.435$\\
			\hline
		\end{tabular}
		\caption{Portion of {the density population in each class} at time $t=200$  in the reference case {with no event (baseline scenario)}, 
			in the slow shock scenario, and the ratio between the two values for the three-classes and five-classes cases.}\label{tab1}
	\end{table}
	
	We observe that, for the three classes case, the most affected class is the lower one. 
	However, the middle class also has its downfall. 
	Indeed, after a time $t=200$, the population of both classes is approximately one-half of what it would have been without the external event. 
	On the other hand, the upper class is less affected by the consequences of the slow shock. 
	This is further stressed in the case $n=5$, where the effects of the slow shock weigh almost exclusively on the lowest class. 
	Indeed, its final value is reduced to a third with respect to the baseline scenario. 
	As for the other classes, the shock has an approximately halving effect for each one. 
	We may infer that the effect of a slow shock needs a certain time to impact each income class, in particular the lower one. 
	In fact, during the initial stage, the solution slightly differs from the one of the baseline scenario; then, after the peak, the effects emerge more clearly. For instance, the impact of climate changes may be modeled by using this slow shock structure, since the event starts at a certain moment, but when they get a certain peak, the effects are clear. In particular, the lower classes pay the toughest consequences. 
	Indeed, lower-income communities often reside in areas subject to environmental risks. 
	Unfortunately,  they may not have the possibility to invest in resilient infrastructure, retrofit homes, or relocate to safer areas.  
	Moreover, lower-income groups are often employed in sectors highly sensitive to climate variations, such as agriculture or informal labor, and often they depend heavily on natural resources for their livelihoods.
	{This tendency is also reflected in the overall decrease in total density. Moreover, we highlight that the crisis amplifies the reduction of the middle-class fraction, leading to a population more polarized between the lower and upper classes.}
	It is also worth observing the reduction of oscillations for the middle class in the case $n=3$ and for lower-middle, middle, and upper-middle in the case $n=5$. 
	In particular, this aspect reflects a reduction in the birth rate for these income classes, revealing the difficulties 
	of these individuals in planning a future with children.

	%	
	%	%%%%%%%%%%%%%%%%
	%	
	\subsubsection{Sudden shock scenario}
	\label{ssecS2}
	
\begin{figure}[ht!]
		\centering	
		\includegraphics[scale=0.143]{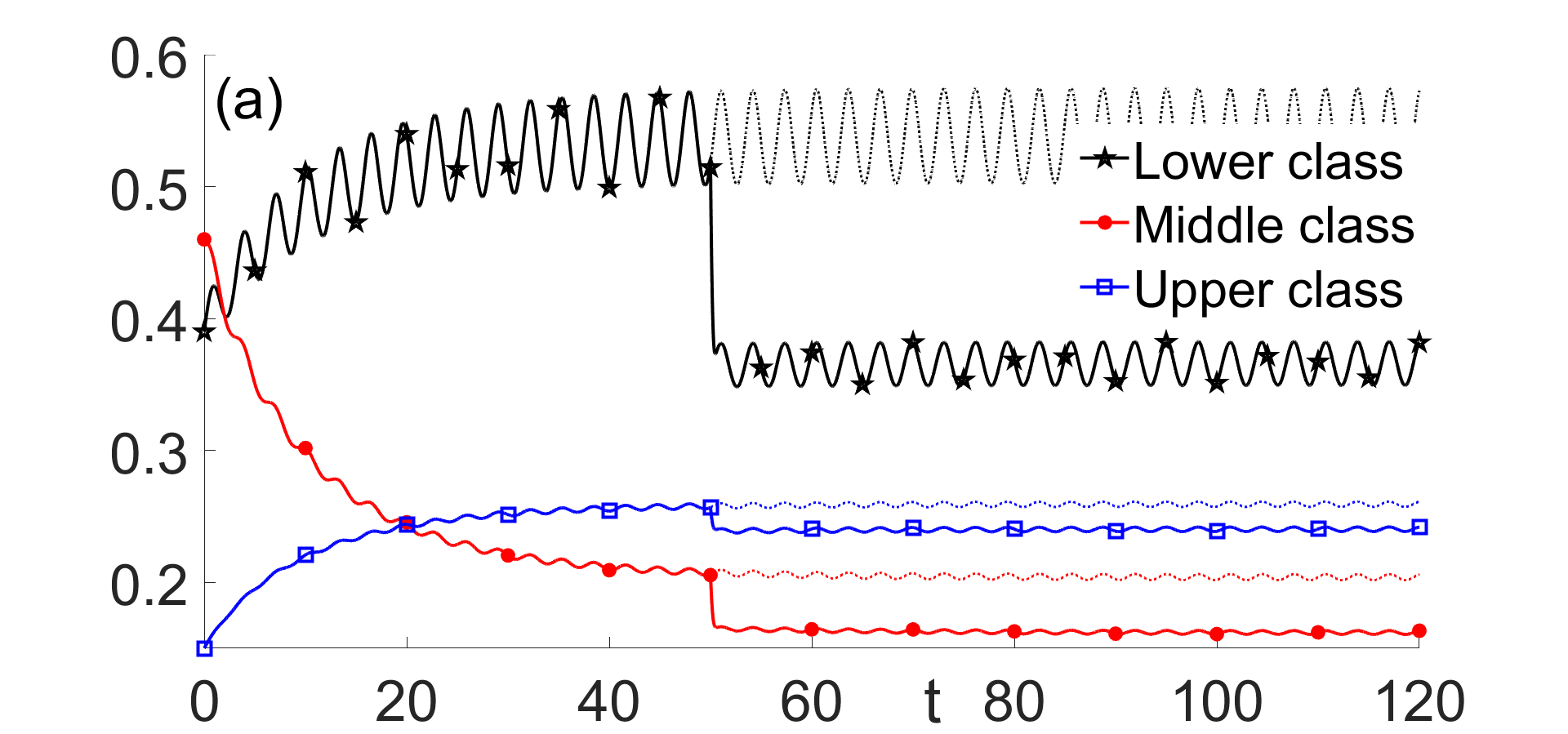}
		\includegraphics[scale=0.143]{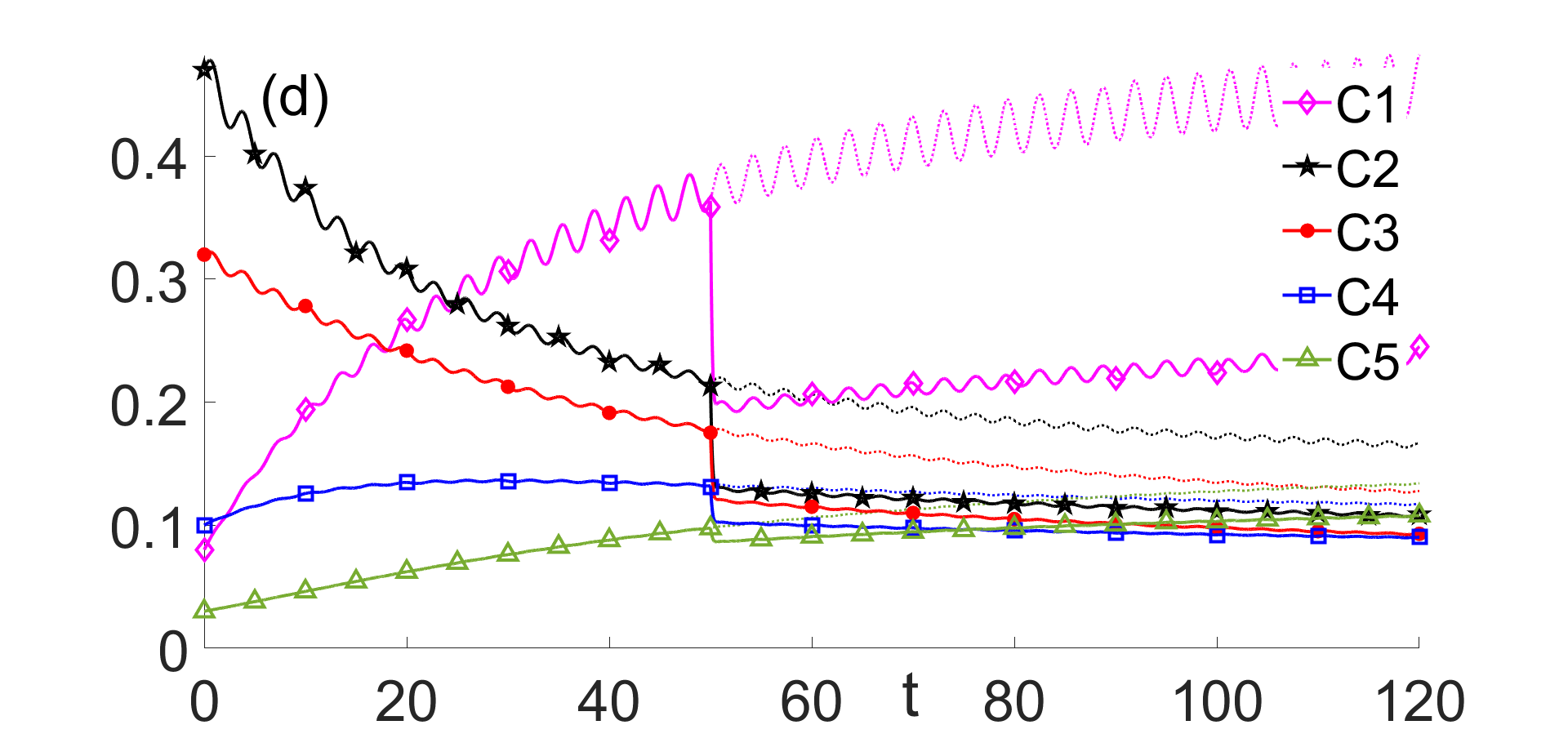}\\
		\includegraphics[scale=0.143]{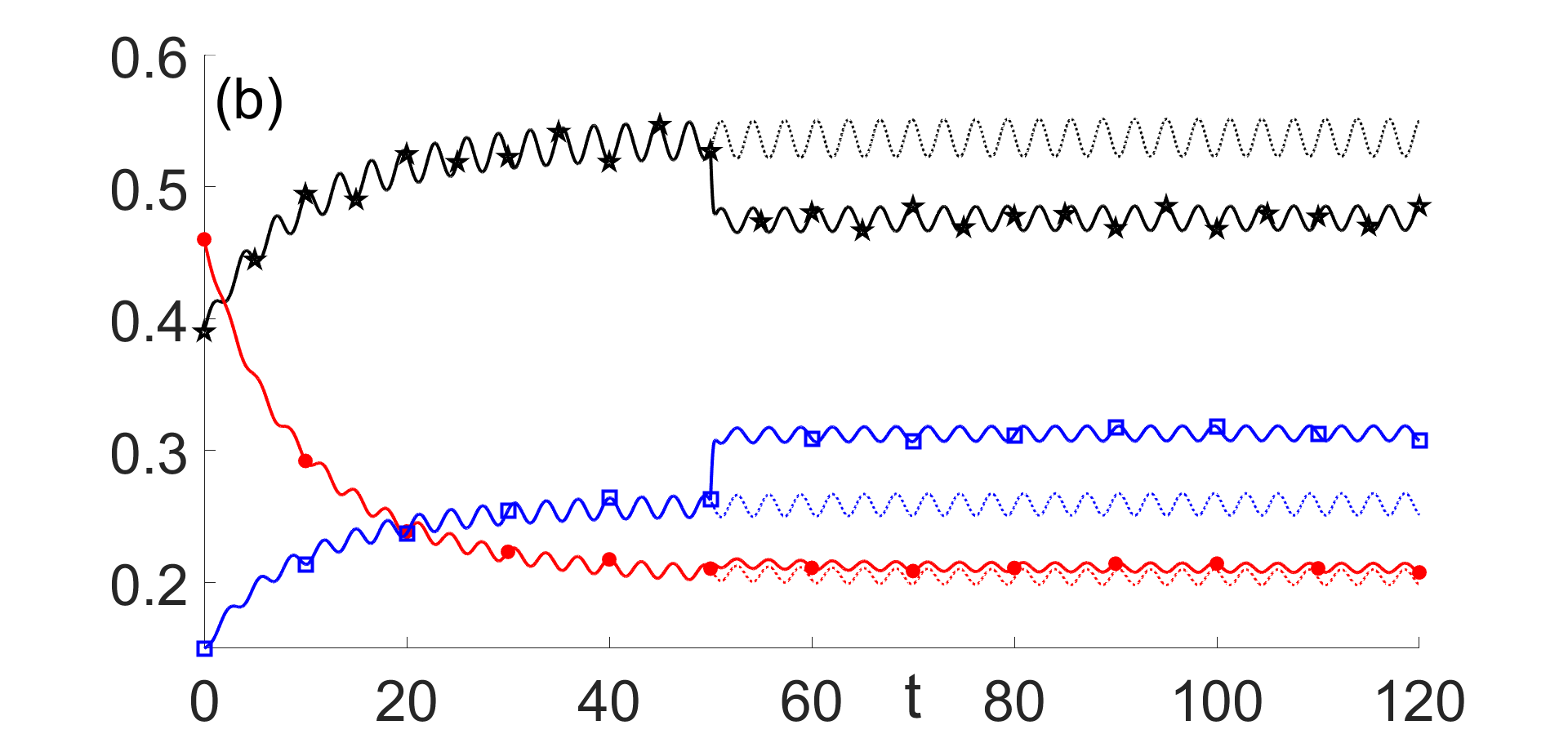}
		\includegraphics[scale=0.143]{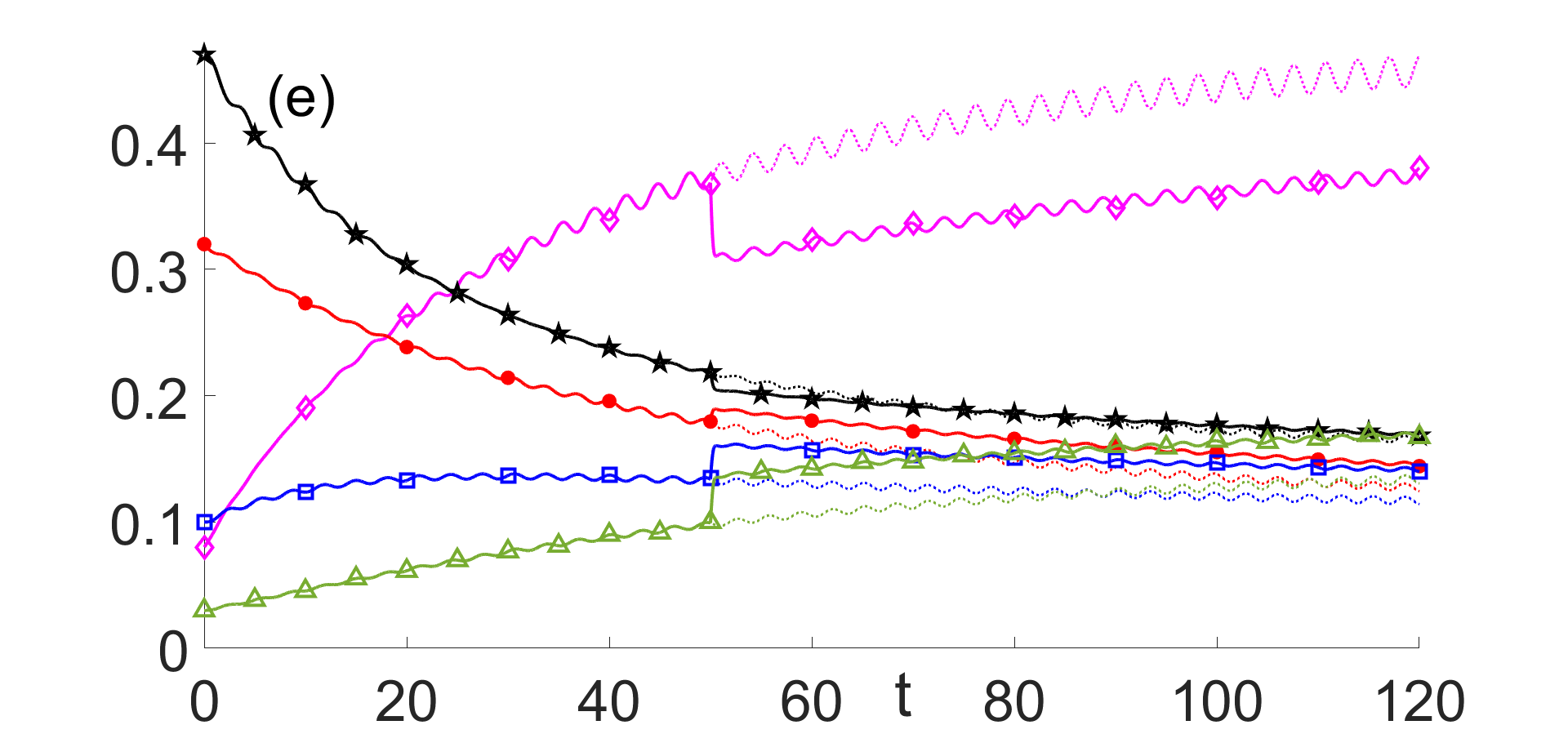}\\
		\includegraphics[scale=0.143]{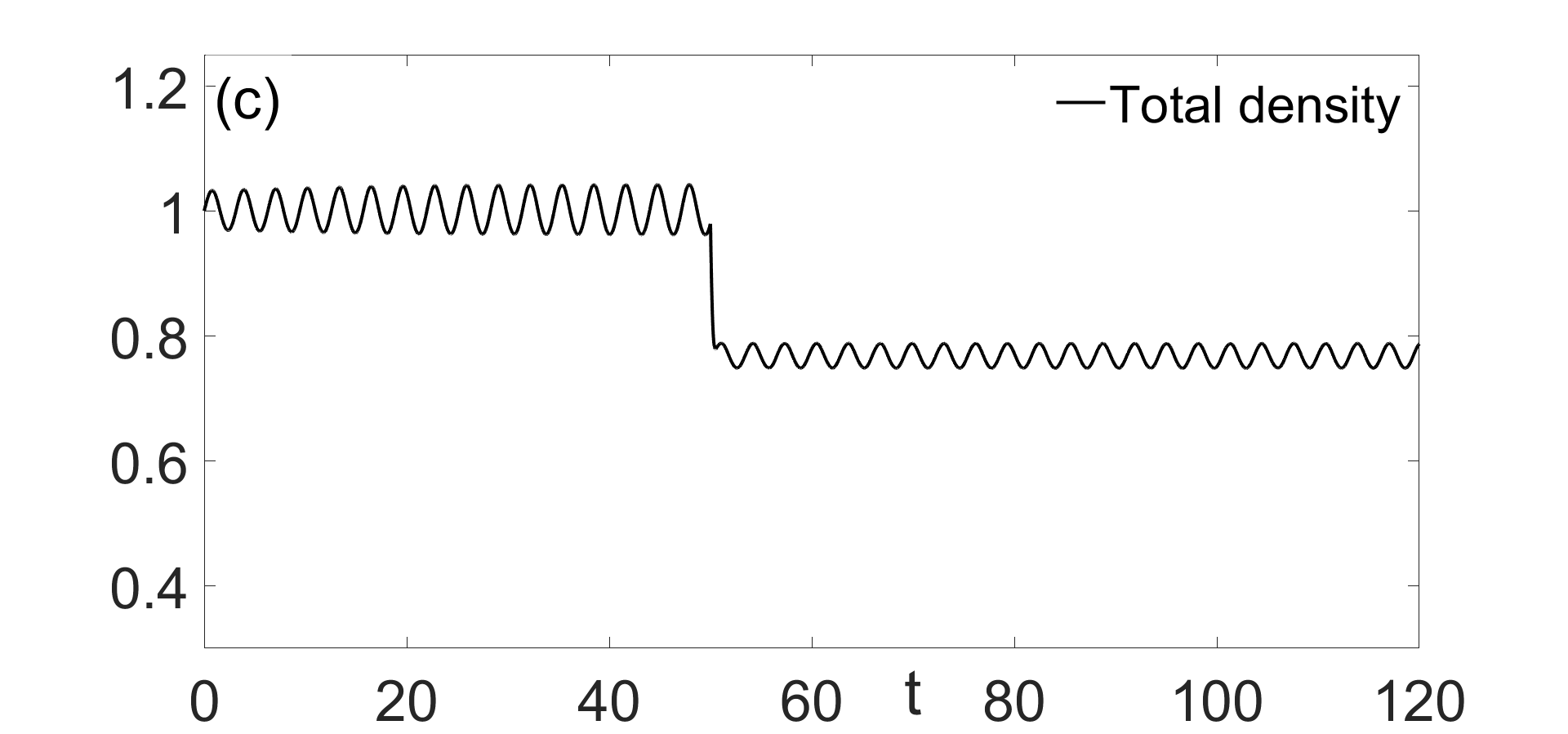}
		\includegraphics[scale=0.143]{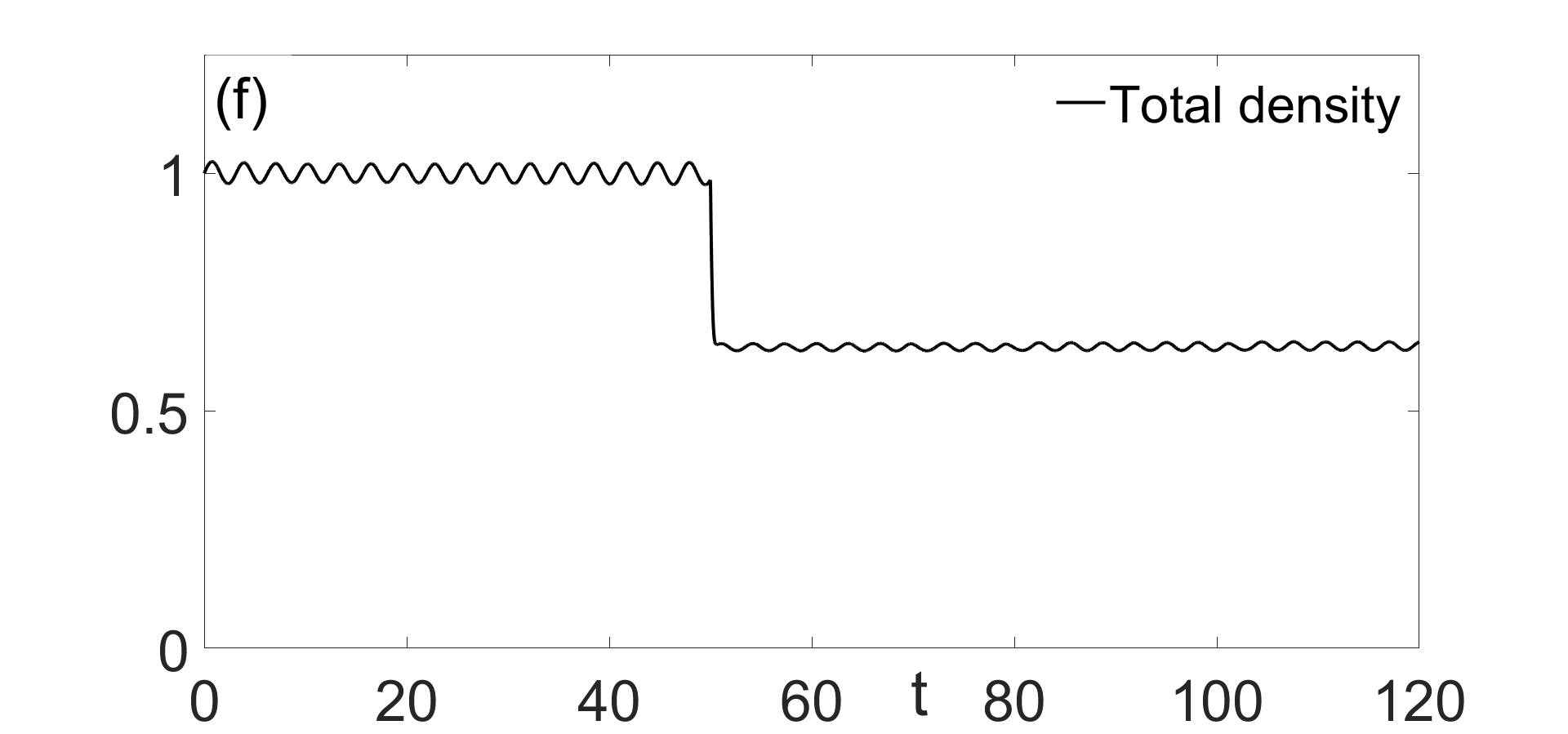}\\
		\caption{\textit{Sudden shock scenario.}
			Time evolution of the population distribution functions when a sudden shock occurs at $t=50$ with $\sigma=0.2$, compared to the reference case (dotted lines). 
			{Left column: three-classes population with parameters as in \eqref{parmetri conservativi}-\eqref{ParsMu1}-\eqref{ParsAl1}
				and initial data \eqref{InData1}. 
				Right column: five-classes population with parameters as in \eqref{parmetri conservativi}-\eqref{ParsMu2}-\eqref{ParsAl2}
				and initial data \eqref{InData2}. 
				Panels {(a) -(d)}: distribution functions; 
				panels {(b) - (e)}: ratio between distribution functions and total density; 
				panels {(c) - (f)}: total density.}}
		\label{Shock}
	\end{figure}

	We now consider the impact of a {sudden shock} that
	is modeled by an external force field with parameters $\lambda_{i}(t)$ of type \eqref{shock2}. We perform {the related simulations} for both cases $n=3$ and $n=5$ in Figure \ref{Shock}, compared to the baseline scenario. Moreover, in Table \ref{tab2} the values of the solution up to time $t=200$, both for sudden shock and baseline scenario, are reported.
	\begin{table}[ht!]
		\centering
		\begin{tabular}{|c||c|c|c|c|}
			\hline
			&Lower&Middle &Upper&Total\\
			\hline
			\hline
			Baseline &$0.507$&$0.202$&$0.257$&$0.966$\\
			\hline
			Sudden shock &$0.348$&$0.16$&$0.239$&$0.747$\\
			\hline
			Ratio &$0.688$&$0.793$&$0.928$&$0.773$\\
			\hline
		\end{tabular}
		\vskip 0.3cm
		\begin{tabular}{|c||c|c|c|c|c|c|}
			\hline
			&C1 & C2 & C3 & C4 & C5&Total\\
			\hline
			\hline
			Baseline &$0.456$&$0.15$&$0.111$&$0.11$&$0.145$&$0.972$\\
			\hline
			Sudden shock&$0.249$&$0.097$&$0.08$&$0.084$&$0.117$&$0.627$\\
			\hline
			Ratio&$0.545$&$0.646$&$0.72$&$0.764$&$0.807$&$0.645$\\
			\hline
		\end{tabular}
		\caption{Portion of the density population in each class at time $t=200$  in the reference case with no event (baseline scenario), 
			in the sudden shock scenario,
			and the ratio between the two values for the three-classes and five-classes cases.}
		\label{tab2}
	\end{table}
	
	The sudden collapse of values for each class emerges in both cases, as expected for such a situation. Nevertheless, the final impact, i.e. at $t=200$, of a scenario with sudden shock is less if compared to the previous case with slow shock, and this is true for both situations. In particular, in the case of $5$ income classes, the lower one starts increasing again soon after the event. {We stress again that the upper classes are those less affected by the {sudden} shock.  Moreover, the oscillatory pattern persists after the shock, also if slightly toned down. An example of sudden shock may be a war, or a particular environmental catastrophe, since their effects on a population appear immediately. Moreover, the lower classes are the ones most exposed to risks and consequences. Therefore, the effects of a sudden shock emerge more quickly if compared to what happens with a slow shock. On the other hand, looking at the final results related to the two scenarios, we observe that, after the same time, the values of functional subsystems are higher in the case of a sudden shock in comparison with the case of a slow shock. This might seem strange, but it is reasonable if we consider that a sudden shock directly appears at its peak compared to a slow shock. However, during a sudden shock, the impact on each functional subsystem is more immediate and leads to a more pronounced effect on each population. Thus we emphasize that the defining characteristic of a sudden shock is its abrupt onset at its peak, which results in the collapse of all income classes, rather than its long-range outcomes.

		\subsubsection{Two successive sudden shocks scenario}
		\label{ssecS4}
		
		Finally, we model a {scenario representing the} effect of two {successive} sudden shocks, occurring some time apart. {In particular, the first one abruptly takes place at time $t=50$ and the second one at $t=90$, in line with the previous case.} To this aim, the external force field writes
		\begin{equation}\label{shock4}
			\lambda_{i}(t)=\alpha_{i}\,\left[\frac{1}{\sigma_1\sqrt{2\pi}}\exp\left(-\frac{(t-\nu_1)^2}{2\sigma_1^2}\right)\mathbf{I}_{\{t\geq\nu_1\}}\,+\frac{1}{\sigma_2\sqrt{2\pi}}\exp\left(-\frac{(t-\nu_2)^2}{2\sigma_2^2}\right)\mathbf{I}_{\{t\geq\nu_2\}}\right],
		\end{equation}
		where, $\nu_1=50$, ${\nu_2=90}$,
		$\sigma_1=\sigma_2=0.2$. {Also in this case, despite it exhibiting two discontinuity points, the locally in-time existence of a positive solution is still ensured.} In Figure \ref{2Shocks} the behavior of the different income classes affected by the two shocks is reported, while the values for each class at time $t=200$ reads, compared to the baseline scenario, in Table \ref{tab4}.
		\begin{figure}[ht!]
			\centering	
		\includegraphics[scale=0.143]{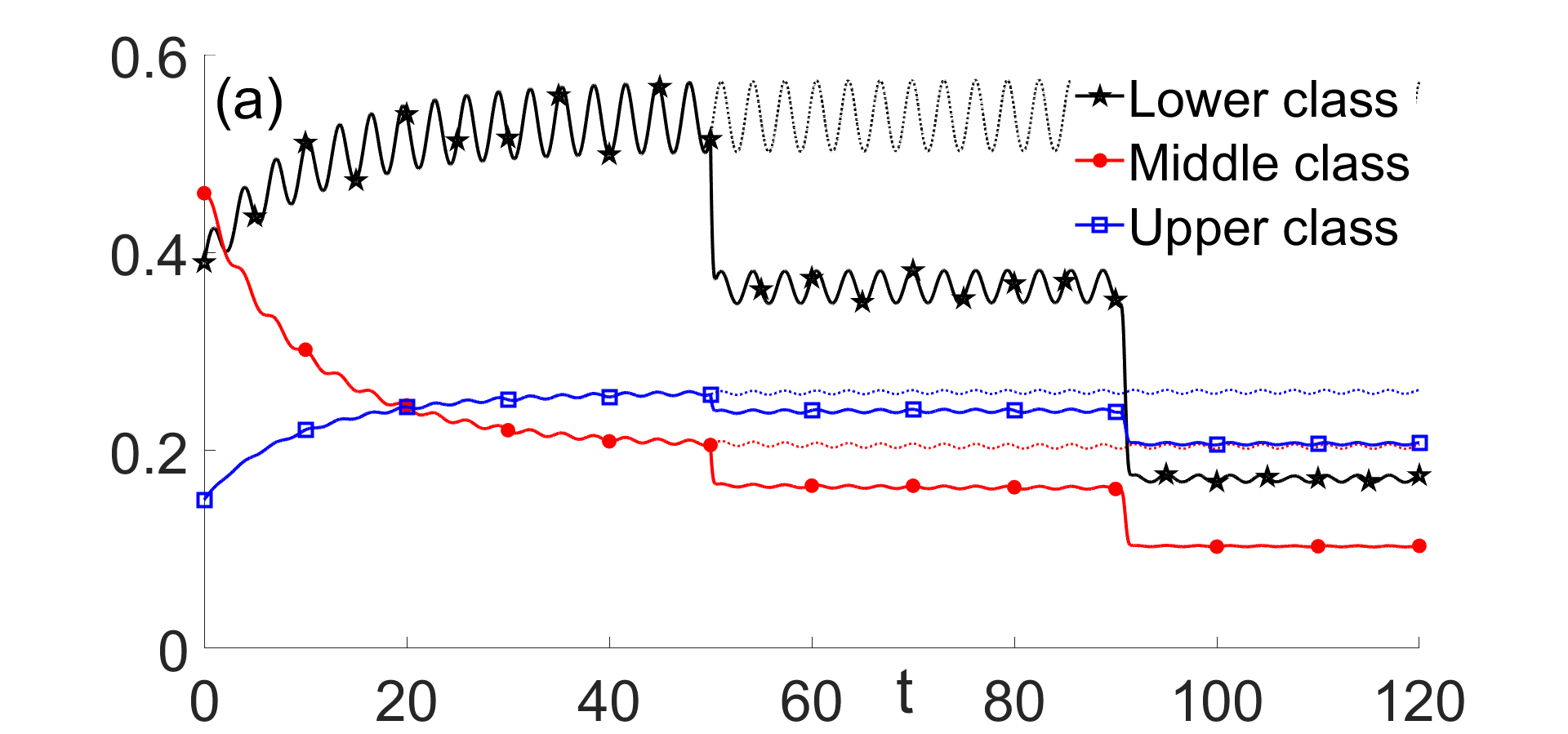}
		\includegraphics[scale=0.143]{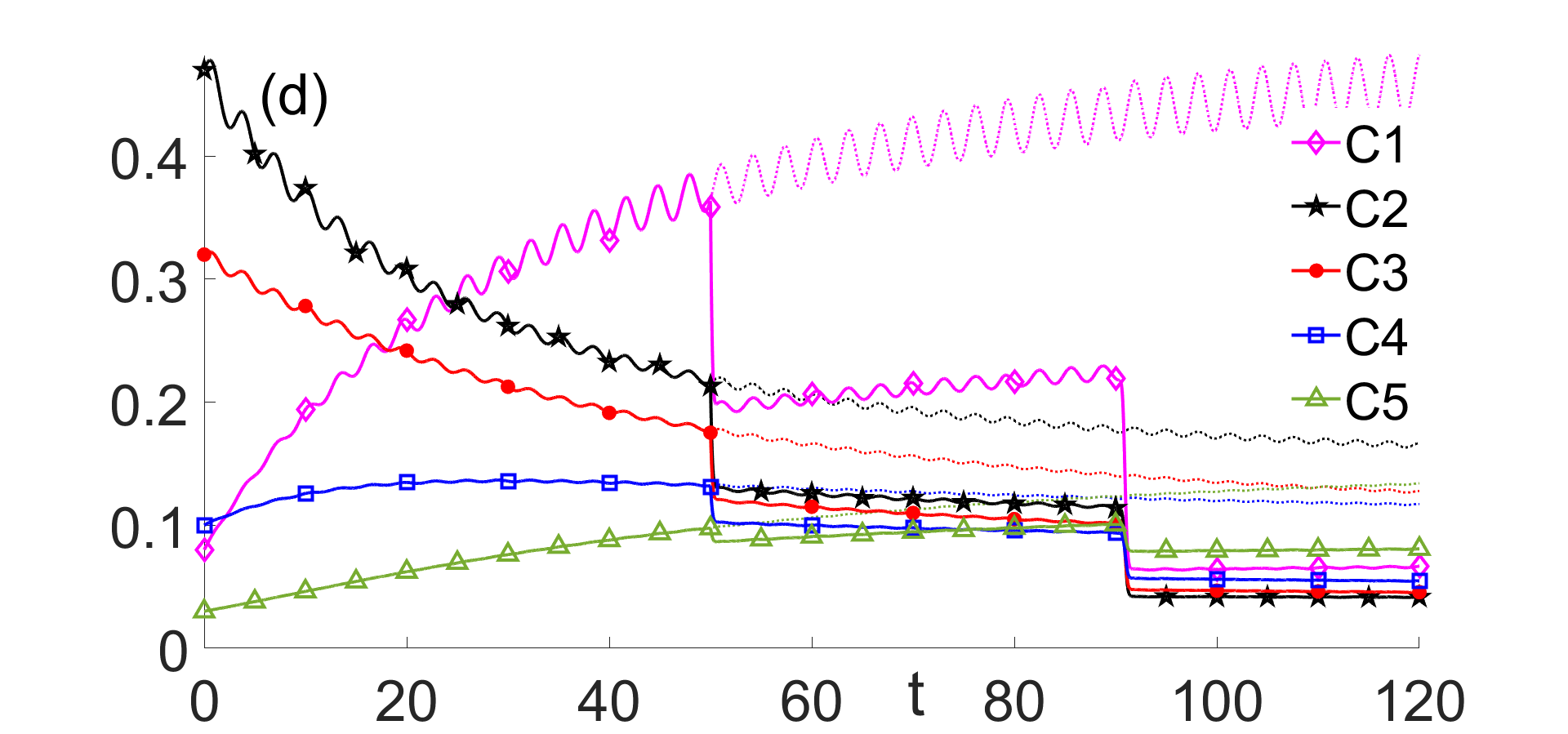}\\
		\includegraphics[scale=0.143]{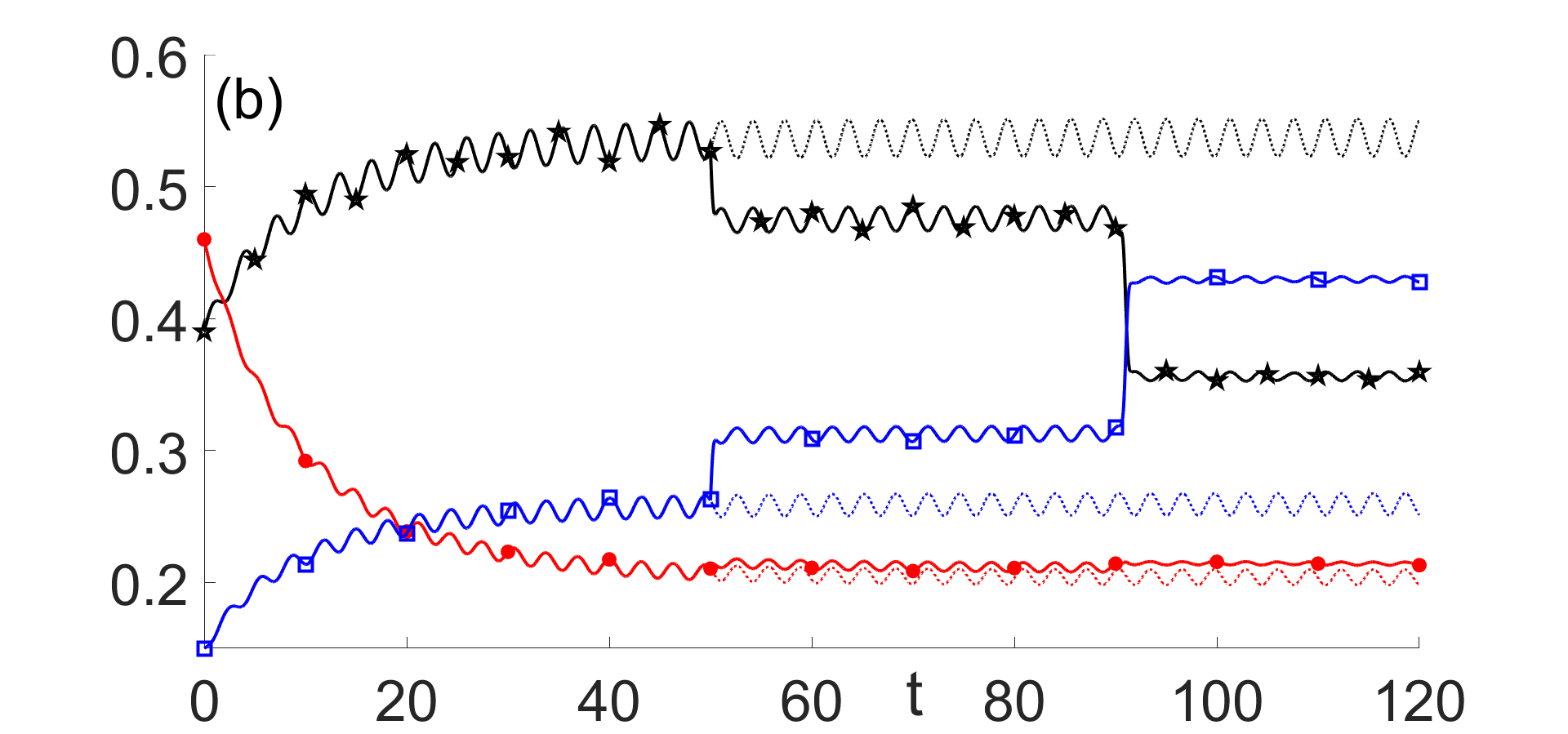}
		\includegraphics[scale=0.143]{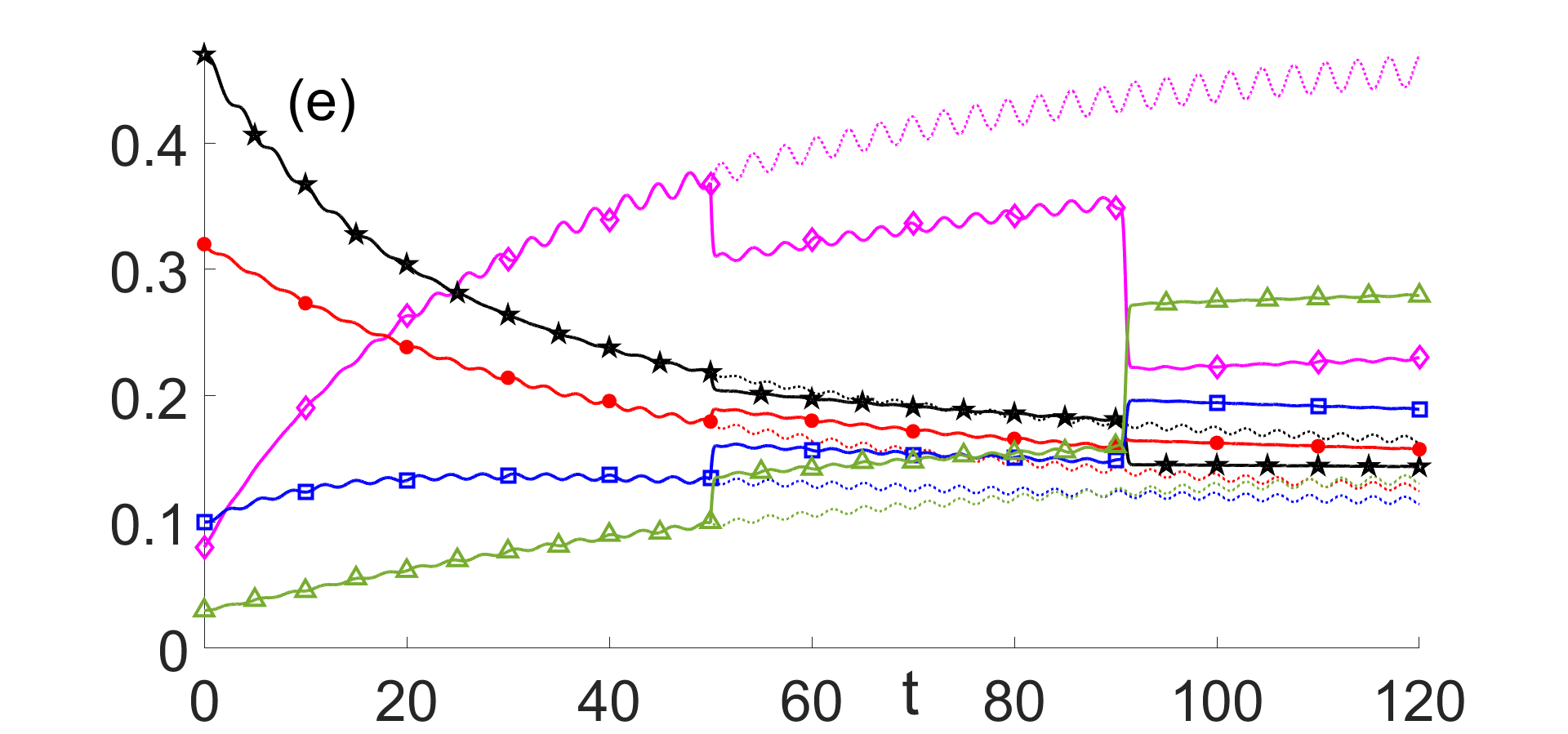}\\
		\includegraphics[scale=0.143]{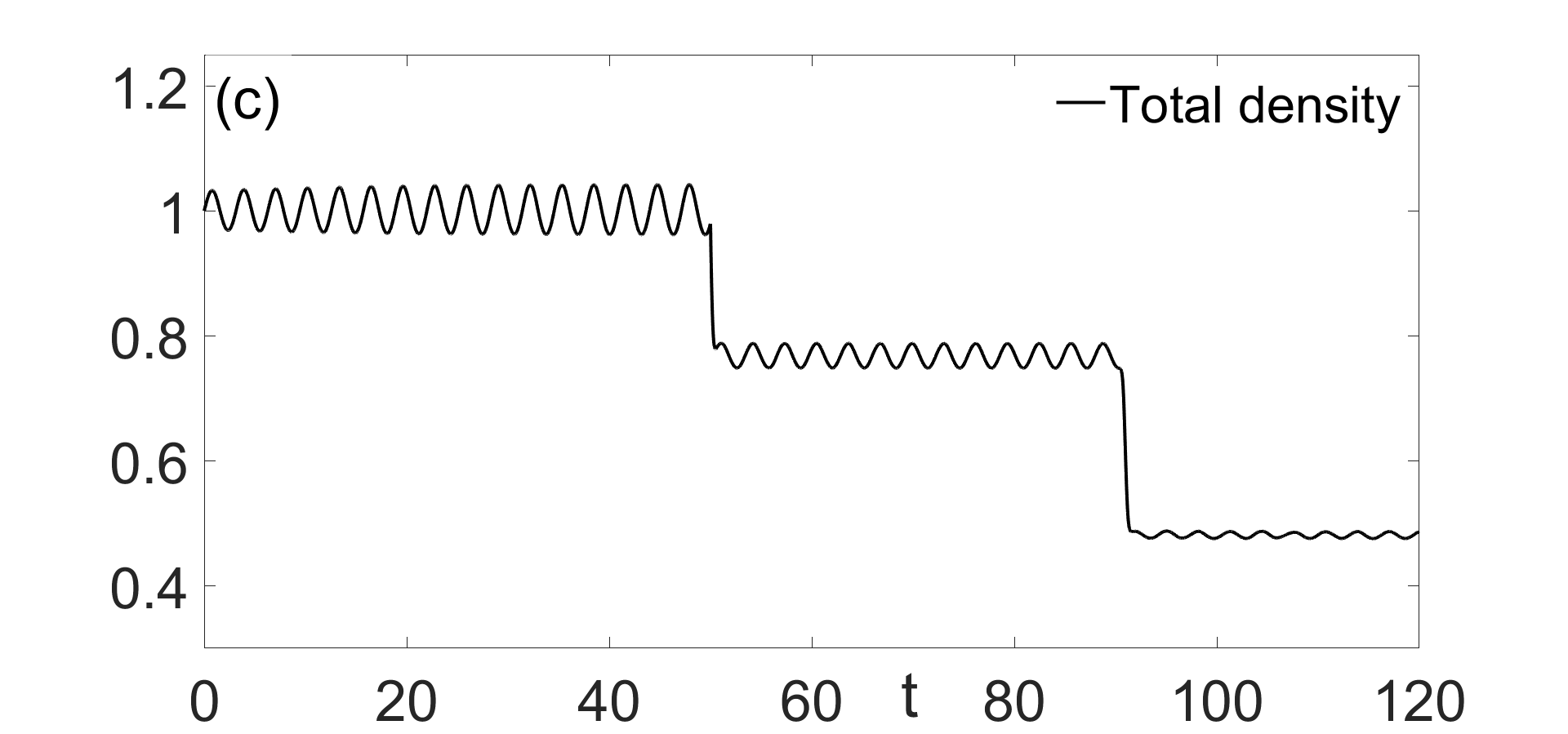}
		\includegraphics[scale=0.143]{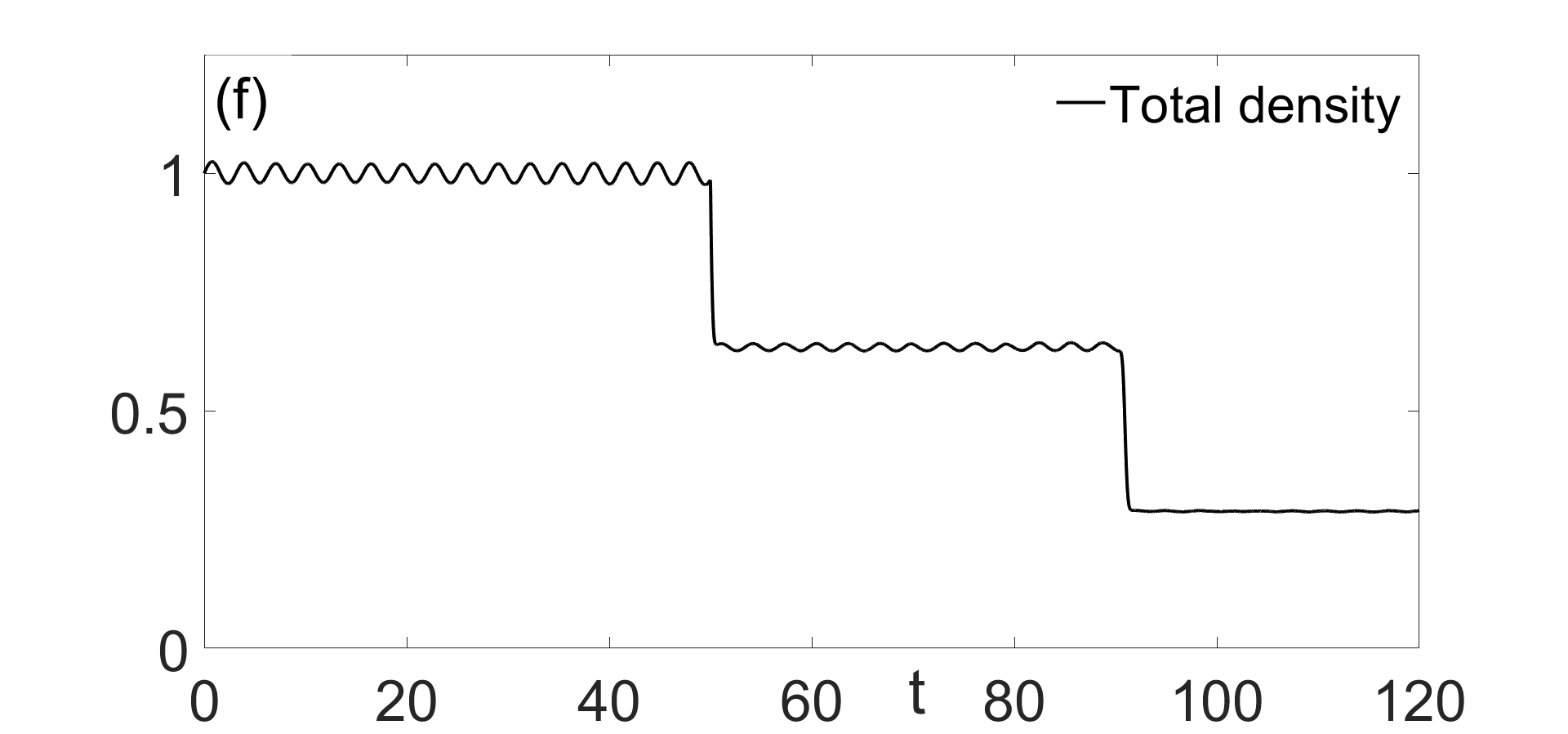}\\
			\caption{\textit{Two successive sudden shocks scenario.}
				Time evolution of the population distribution functions when a sudden shock occurs at $t=50$ with $\sigma_1=0.2$, followed by another one at $t=90$ with $\sigma_1=\sigma_2=0.2$, compared to the reference case (dotted lines). 
				{Left column: three-classes population with parameters as in \eqref{parmetri conservativi}-\eqref{ParsMu1}-\eqref{ParsAl1}
					and initial data \eqref{InData1}. 
					Right column: five-classes population with parameters as in \eqref{parmetri conservativi}-\eqref{ParsMu2}-\eqref{ParsAl2}
					and initial data \eqref{InData2}. 
					Panels {(a) - (d)}: distribution functions; 
				        panels {(b) - (e)}: ratio between distribution functions and total density; 
				        panels {(c) - (f)}: total density.}}
			\label{2Shocks}
		\end{figure}
		
		\begin{table}[ht!]
			\centering
			\begin{tabular}{|c||c|c|c|c|}
				\hline
				&Lower&Middle &Upper&Total\\
				\hline
				\hline
				Baseline&$0.507$&$0.202$&$0.257$&$0.966$\\
				\hline
				Two shocks&$0.168$&$0.103$&$0.206$&$0.477$\\
				\hline
				Ratio&$0.332$&$0.508$&$0.8$&$0.494$\\
				\hline
			\end{tabular}
			\vskip 0.3cm
			\begin{tabular}{|c||c|c|c|c|c|c|}
				\hline
				&C1 & C2 & C3 & C4 & C5&Total\\
				\hline
				\hline
				Baseline&$0.456$&$0.15$&$0.111$&$0.11$&$0.145$&$0.972$\\
				\hline
				Two shocks&$0.071$&$0.04$&$0.042$&$0.051$&$0.085$&$0.289$\\
				\hline
				Ratio&$0.156$&$0.267$&$0.374$&$0.467$&$0.585$&$0.297$\\
				\hline
			\end{tabular}
			\caption{Portion of the density population in each class at time $t=200$  in the reference case with no event (baseline scenario),
				in scenario 4 of two sudden shocks, and the ratio between the two values for the three-classes and five-classes cases.}
			\label{tab4}
		\end{table}
		
		We report that the outcomes are very close to the previous case with sudden and slow shock, i.e. the impact of the second shock enforces the decreasing effect of the first one, most evidently for lower classes, both for case $n=3$ and $n=5$. We may conclude that the presence of a second shock following a first sudden shock impacts a society, regardless of its behavior in time. This can be justified by the fact that the classes, from lower to middle, have already been affected by the effects of the first shock. Therefore, the second shock, even if slow, has the same impact on the overall dynamics.}
	
	When we compare the last two scenarios, where two consecutive shocks occur, with the previous ones involving a single shock, some observations can be made. In both cases, for $n=3$ and $n=5$, the main impact still affects the lower class, resulting in a significant population decline. However, a peculiar phenomenon emerges when the second slow or sudden shock occurs. The upper class overtakes the lower classes shortly after the peak of the second slow shock or immediately when the second sudden shock occurs. This suggests that a second shock turns out to be unsustainable for the lower classes, which have already been severely tried by the first one. Whereas, only the upper classes can stand this further situation.
	It is worth noting that in the case of $n=3$, the middle class definitely collapses, and the oscillations are even smaller compared to the scenario with a single shock. 
	
	This outcome is consistent with the real-world experience, as birth rates often decrease following shock events due to economic issues.
	For example, we can imagine a real-life situation with two shocks, such as a natural disaster followed by an employment crisis. Both of these events represent sudden shocks affecting all income classes within society. Typically, lower-income individuals are most affected by natural disasters, especially those whose homes are not structurally adequate to withstand them. 
	%Additionally, the second shock, such as an epidemic outbreak, further impacts those who were already struggling due to the first shock, potentially because they live in temporary, overcrowded housing with lacking sanitation and limited access to healthcare. 
	Consequently, the upper class, having been less affected by the shocks, becomes the more populated segment of society.
	
	\medskip
	{At this point, we aim to analyze the behavior, in all considered cases, 
	of the total income $\overline\xi(t)$ and of the mean wealth, that we define as 
	$$
	\omega(t)=\frac{\overline\xi(t)}{\rho(t)\,r_n}.
	$$
	The dynamics of these quantities is illustrated in Figure \ref{Total}.
	\begin{figure}[ht!]
		\centering	
		\includegraphics[scale=0.143]{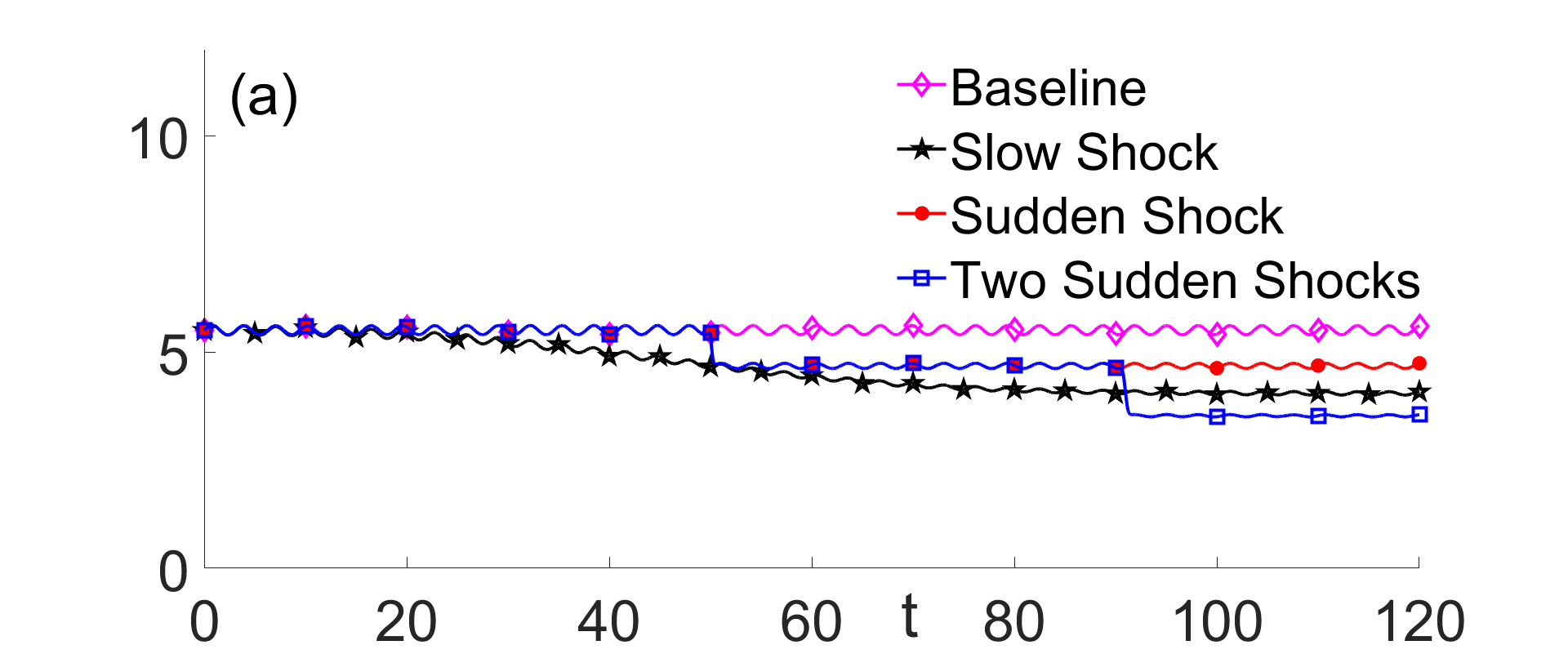}
		\includegraphics[scale=0.143]{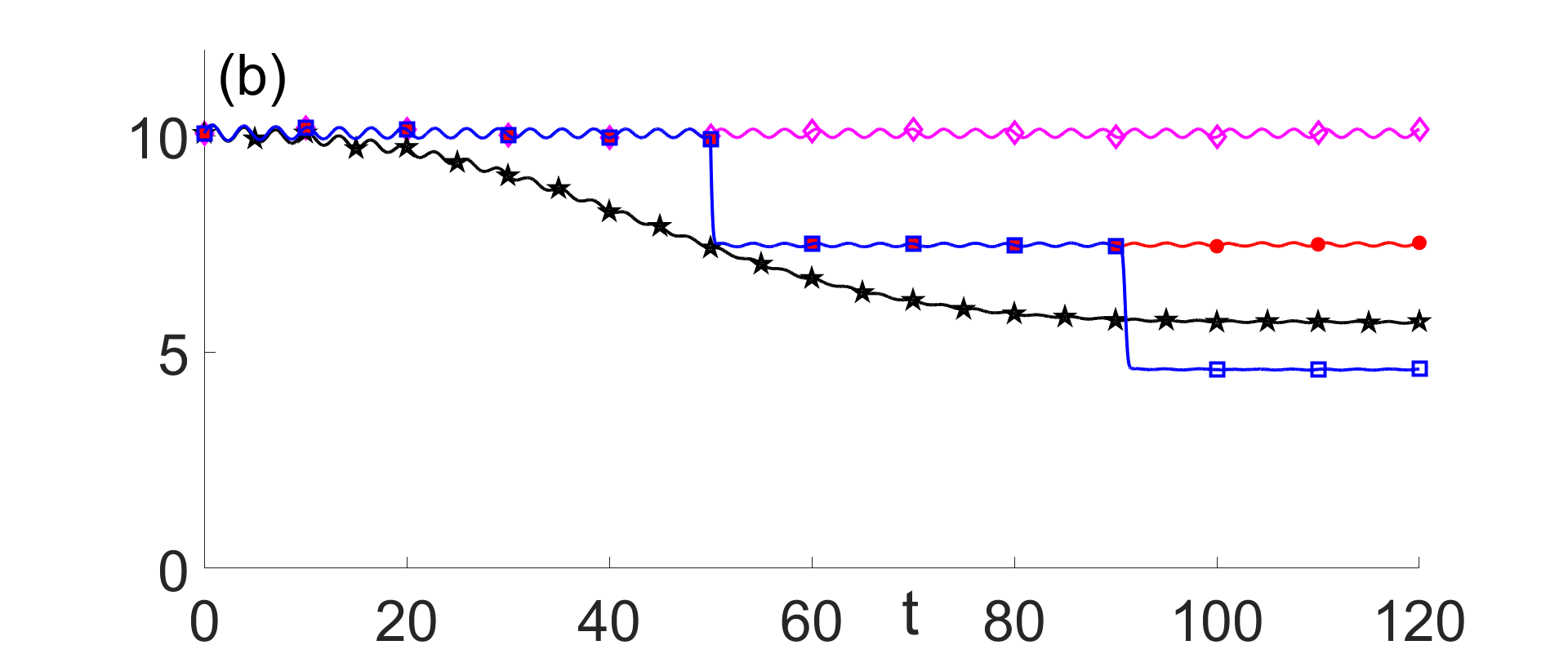}\\
		\includegraphics[scale=0.143]{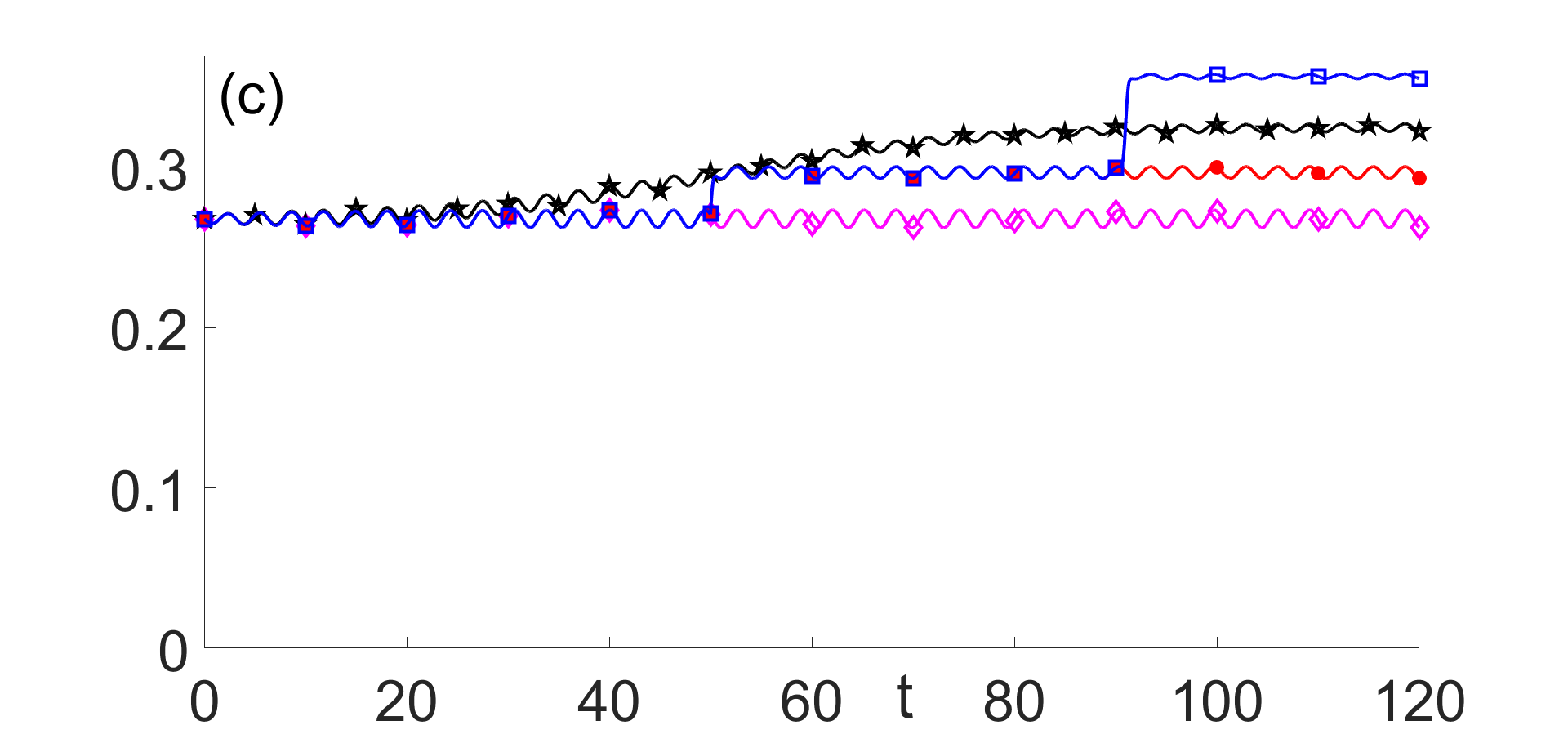}
		\includegraphics[scale=0.143]{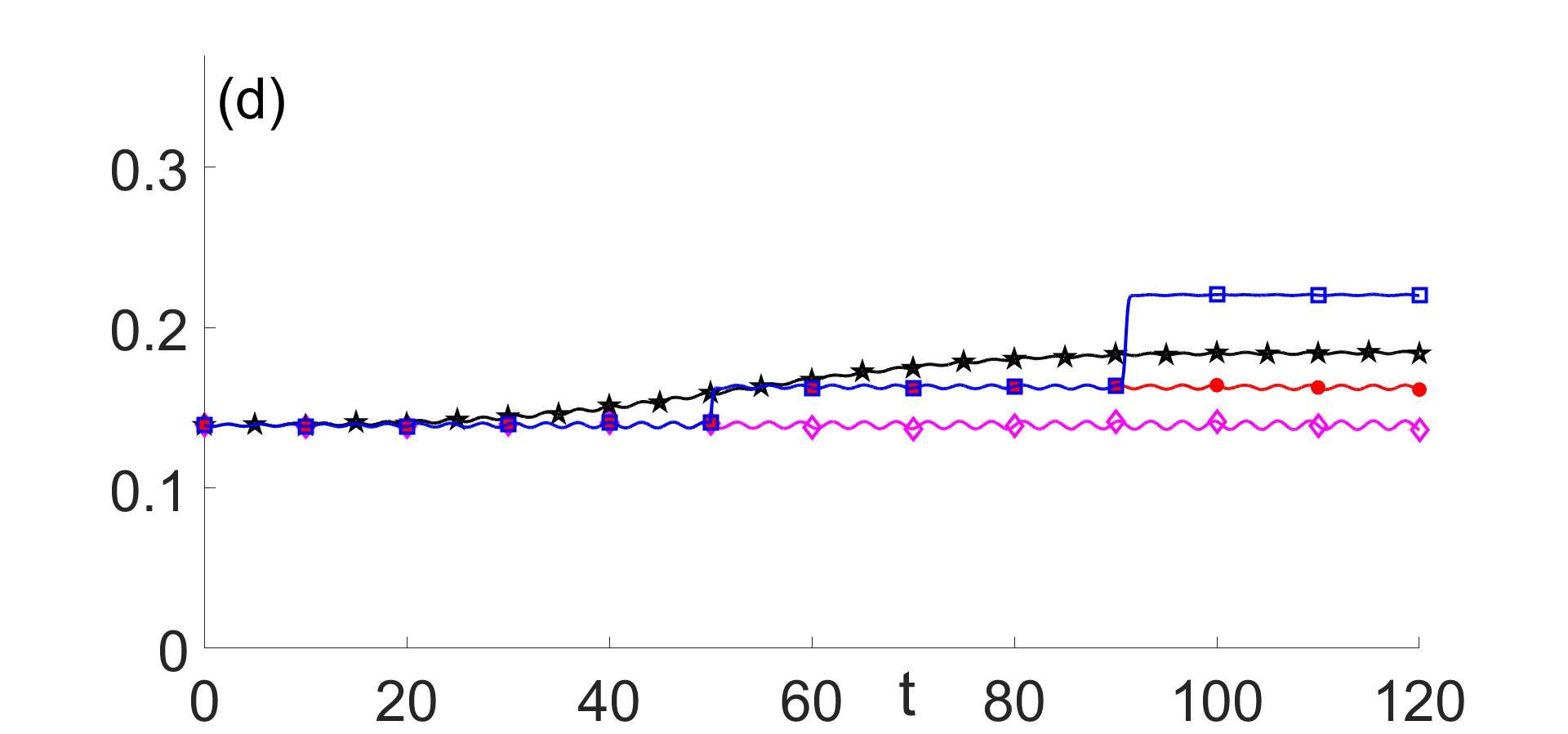}\\
		\caption{\textit{Total income and mean wealth.}
		First row: behavior of total income for the cases of $n=3$ (panel (a)) and $n=5$  (panel (b)) in the four considered scenarios. Second row: behavior of mean wealth for the cases of $n=3$ (panel (c)) and $n=5$  (panel (d)) in the four considered scenarios.}
		\label{Total}
	\end{figure}
	
	In panels (a) and (b) the total income for the cases of $n=3$ and $n=5$ wealth classes, respectively, is reported, comparing the trends for each one of the scenarios previously modeled. We may observe that, as during crisis and shocks the total density of the population decreases, the total income is lowered as well. 
	In panels (c) and (d), instead, we report the mean wealth  for the cases of $n=3$ and $n=5$ wealth classes. In this context, we observe that crises and shocks lead to an increase in the average wealth. This effect is primarily due to the fact that the lower classes are more severely impacted by external events and, as previously discussed, the middle class tends to vanish. In other words, the wealthier fraction of the population is the one that survives.
}

		\section{Application to a real scenario: the Hurricane Katrina case}
		\label{secS5}
		In this section, we adapt the kinetic framework described in Section \ref{secmark} to a real example of a shock event impacting a certain population. More precisely, we consider the impact of Hurricane Katrina which made landfall in southeast Louisiana and southwest Mississippi on August 29, 2005. The majority of the damage was represented by the flooding that, after the hurricane, impacted the city of New Orleans MSA, and its surrounding area. The data documented in the literature (see datasets reported in \cite{shaughnessy2010income}) show the sharp rise in the housing vacancy rate and the drastic drop in the population of New Orleans from September to December 2005.
		Economists are concerned with the widespread effects of natural disasters on local economies. These events cause physical damage, disrupt resource supplies, and affect households, leading to labor interruptions. Assessing both the short-term and long-term economic impacts is challenging due to limited data, making it a complex area of study (see \cite{shaughnessy2010income} and references therein). For the particular case of Hurricane Katrina, works as \cite{landry2007going} have been focused on analyzing how the large-scale evacuation before the hurricane influenced the distribution of income. The ability of a household to evacuate, return, rebuild, and find employment, in fact, likely depends on its wealth or income. Consequently, the varying impacts on households of different income levels contribute to notable changes in the political, social, and cultural landscape of the city. In paper \cite{shaughnessy2010income}, the authors collect data relevant to the income distribution of New Orleans before the hurricane, from January to August 2005, in the short-term after the hurricane, from September to December 2005, and in the long-term, from August 2005 to December 2007. The authors also compare data for the last two periods to the income distribution of the total U.S. population. Their scope is to depict, using theoretical models based on diverse fitting density functions, how short-run changes can offer valuable early insights into the long-term effects on income following a disaster. They suggest that understanding these immediate changes is crucial for researchers, as it can inform and potentially alter assumptions about economic behavior post-disaster. Additionally, the authors emphasize that permanent shifts in income distribution could lead to different long-term recovery outcomes compared to mere changes in income levels. 
		
		The particular case study presented above provides a suitable framework for applying our model. The short-term effect of Hurricane Katrina can be seen as the impact of a sudden shock on the population, while the long-term effect can be interpreted as a second slower shock.  
		
		First, we adapt the data reported in \cite{shaughnessy2010income}, where the total population is divided into ten income classes, to fit our modeling. To achieve this, we merge the classes with similar data as follows. 
		\begin{itemize}
			\item[] $C1$: Less than $\$ 14.999$,
			\item[] $C2$:  $\$ 15.000$ to $\$ 34.999$,
			\item[] $C3$:  $\$ 35.000$ to $\$ 74.999$,
			\item[] $C4$:  $\$ 75.000$ to $\$ 149.999$,
			\item[] $C5$:  $\$ 150.000$ or more.
		\end{itemize}
		
		At this point, we want to focus on the main feature of the model, which is the impact of the external force field. For this reason, at this stage, we do not consider the birth and the death terms.
		
\begin{figure}[ht!]
			\centering
			\includegraphics[scale=0.25]{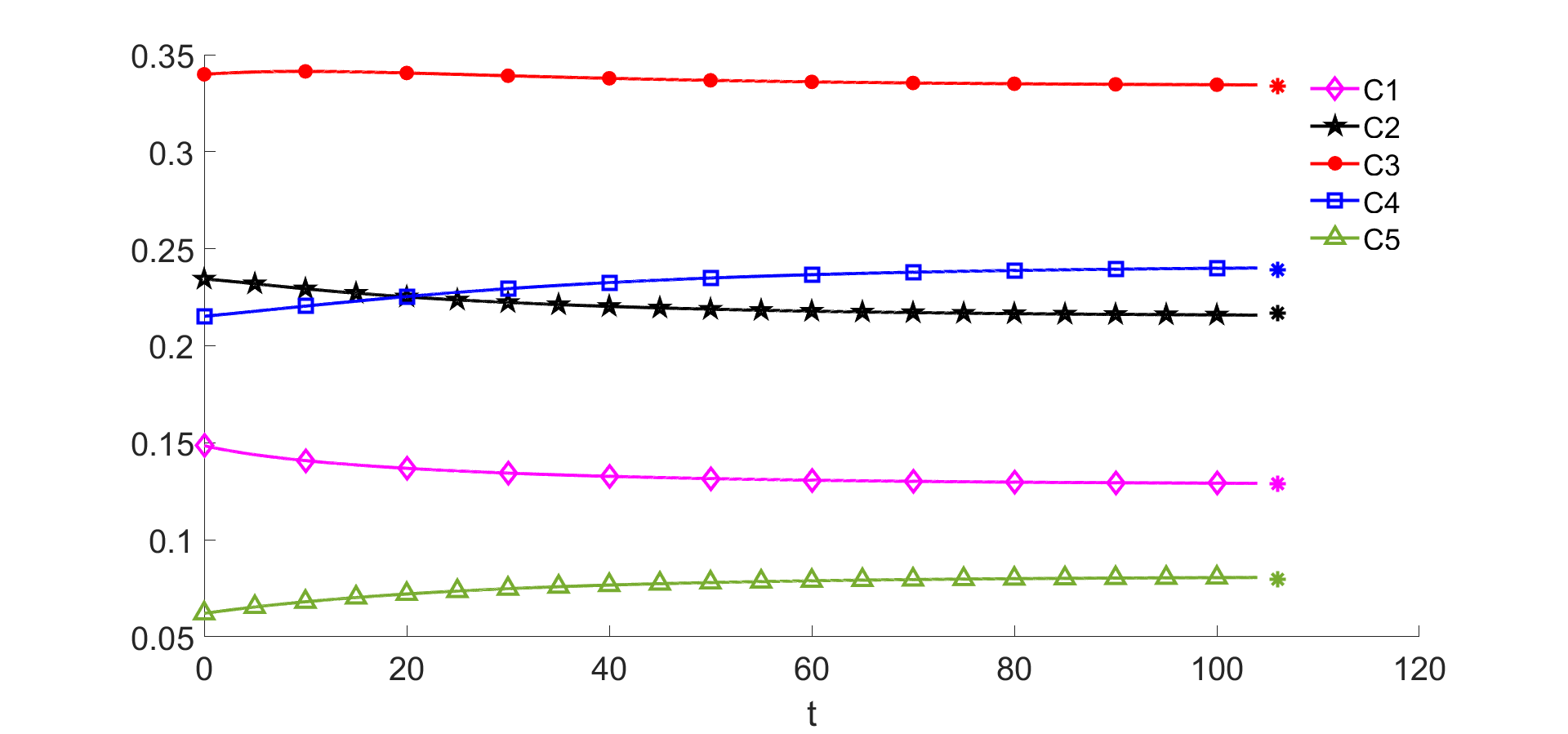}
			\caption{
				Behavior of system \eqref{eqkin}
				without the non-conservative terms, taking as initial data the fractions income distribution of United States at the beginning of 2005 and comparing the final values at the end of 2007 as reported in \cite{shaughnessy2010income}. Interaction frequencies are as in \eqref{RatesAlKat1}.
				The time $t$ is in weeks.}
			\label{NoEventKat}
		\end{figure}

		Let us now build the conservative dynamics. In this case, we want to choose transition probabilities $B_{hk}^i$ directly to reproduce real data. We assume that the dynamics of the New Orleans population (without the shock event) are analogous to the one of the whole US  population reported in \cite{shaughnessy2010income}. 
		Thus, we suppose that interactions occur only among individuals of the same class or belonging to two adjacent classes. We take:
		\begin{itemize}
			\item[--] probability that, after the interaction, the individual belonging to the $i-th$ class passes to the $(i-1)-th$ class (or, if $i=1$, remains in the same class) to be $0.05$;
			\item[--] probability that, after interaction, the individual remains in the same class to be $0.92$;
			\item[--] probability that, after the interaction, the individual belonging to the $i-th$ class passes to the $(i+1)-th$ class (or, if $i=5$, remains in the same class) to be $0.03$.
		\end{itemize}
		Moreover, interaction rates are chosen to reproduce real data, thus:
\begin{equation}
\begin{aligned}
&\eta_{11}=\eta_{21}=\eta_{23}=5\times10^{-4},\, 
\eta_{12}=9.75\times10^{-2},\, 
\eta_{22}=3.4\times10^{-2},
\\
&
\eta_{32}=3\times10^{-3},\,
\eta_{34}=\eta_{43}=2.5\times10^{-3},\, 
\eta_{33}=\eta_{44}=5\times10^{-2},\, 
\\
&
\eta_{45}=4.7\times10^{-3}, \, \eta_{54}=4.5\times10^{-3},\, 
\eta_{55}=4\times10^{-2},
\end{aligned}
\label{RatesAlKat1}
\end{equation}
		with all the remaining interaction rates being equal to zero. 
		{Our current aim is to model directly the fraction of households with respect to the total population, temporarily neglecting the actual number of households within each income class. This {approach} allows us to focus on the relative dynamics of income classes without being influenced by changes in the total number of households, highlighting structural trends in the distribution.}
		Figure \ref{NoEventKat} {shows} the behavior of system \eqref{eqkin}
		without nonconservative terms, taking as initial data the fractions income distribution of US at the beginning of 2005 until 2007.
		The time $t$ is given in weeks.
		Moreover, we compare our final results with the available {data} and mark this data with asterisks. 
		We neglect here the change in the total population, during the two years.
		
		{We emphasize that the real-world scenario depicted in Figure~\ref{NoEventKat} differs substantially from those examined in the previous sections. In particular, while the size of the middle class remains approximately constant over time, the number of households in the lower income classes shows a decreasing trend. This contrast arises from the different modeling assumptions. {In} the earlier sections, we considered a setting where all interaction rates were uniformly set to one, aiming to illustrate the long-term behavior of the model. In the current case, however, the interaction frequencies are calibrated to reflect empirical data and to show how the changes in population fractions are not primarily driven by class-to-class exchanges, but rather emerge as a direct consequence of external events such as shocks and crises.}
		
\begin{figure}[ht!]
			\centering
			\includegraphics[scale=0.25]{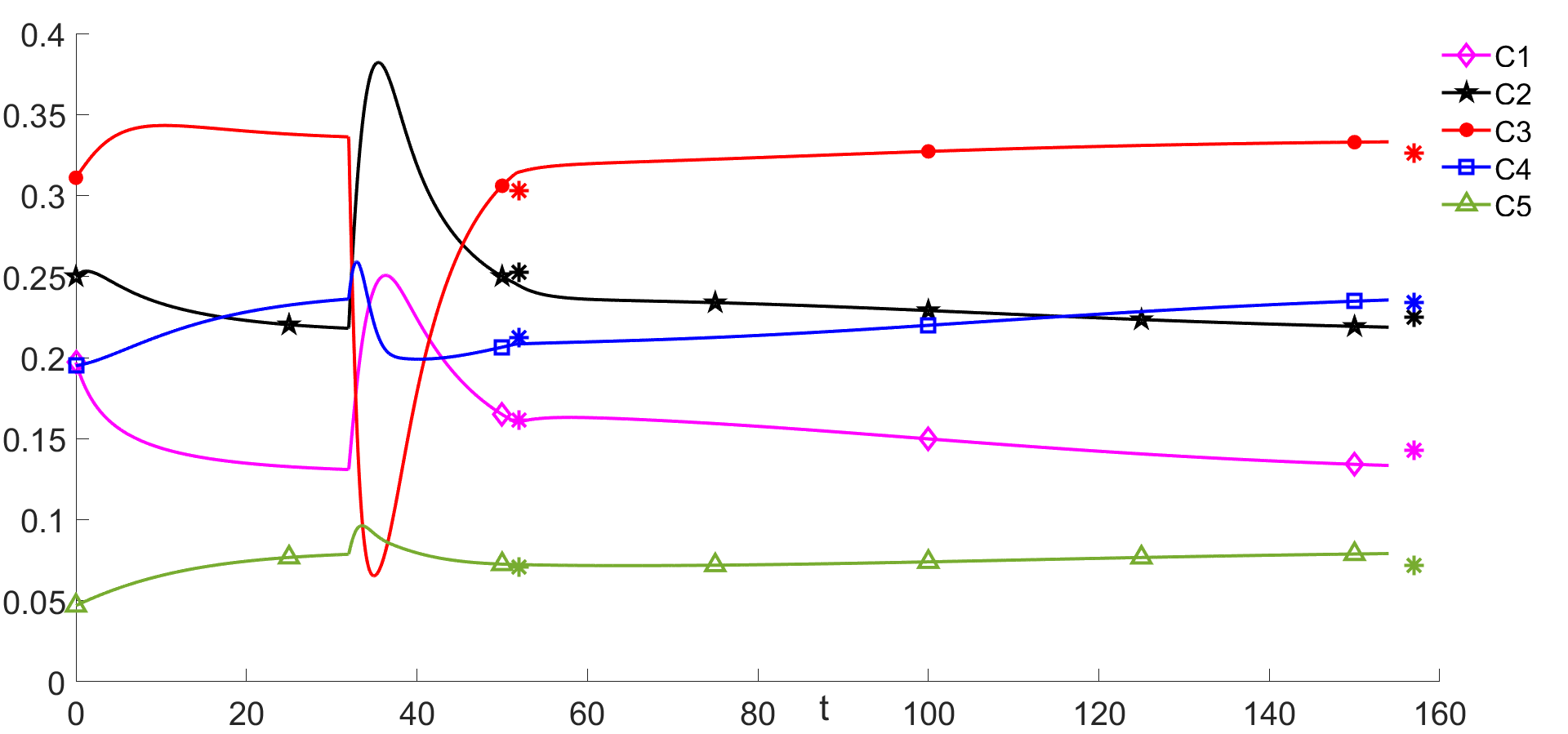}
			\caption{ Behavior of the population after the impact of Hurricane Katrina and the consequent flooding, 
				followed by the related crisis modeled by system \eqref{eqkin}
				{wi}thout the birth/death terms and taking two consecutive shocks modeled by \eqref{shock4}, with $\nu_1=33, \sigma_1=2,\nu_2=52,\sigma_2=50$, taking as initial data the fractions income distribution of New Orleans at the beginning of 2005 \cite{shaughnessy2010income}, interaction frequencies as in \eqref{RatesAlKat1} and parameters as in \eqref{ParsAlKat1}-\eqref{ParsAlKat2}.
				{The} time $t$ is in weeks. }
			\label{ShockKat}
		\end{figure}

		Now, we consider the shock dynamics, in order to reproduce data on the income distribution from the beginning of 2005 to the end of 2007. To model the direct impact of the hurricane and the consequent flooding, we consider then a shock of shape given in \eqref{shock2}, with $\nu=33$ and $\sigma=2$. For the coefficients $\alpha_i$ we take		
		\begin{equation}
			\label{ParsAlKat1}  
			\alpha_1=0.06,\quad \alpha_2=0.12,\quad \alpha_3=-1.25,\quad \alpha_4=-0.2,\quad \alpha_5=-0.1.
		\end{equation}
		On the other hand, the crisis deriving from the difficulty of inhabitants to come back to their houses and find employment during the following two years can be represented by a slow shock still of the form \eqref{shock2},  with $\nu=52$ and $\sigma=50$. In this case, instead, coefficients $\alpha_i$  are
		\begin{equation}
			\label{ParsAlKat2}  
			\alpha_1=0.8,\quad \alpha_2=0.03,\quad \alpha_3= -0.07,\quad \alpha_4=-0.005,\quad \alpha_5=0.1.
		\end{equation}
		{
		We observe that, unlike in the previous generic simulations, the coefficients $\alpha_i$ can be either positive or negative. This results from our choice to model the fraction of households relative to the total population. After the shock, many individuals from upper income classes tend to relocate, leading to negative coefficients for those classes. In contrast, individuals in lower income classes are generally unable to move, so the fraction they represent increases, resulting in positive coefficients.
		A similar pattern is seen in the slower shock scenario, with the difference that reconstruction efforts and the gradual return of residents are also considered.}
		Thus, we perform numerical simulations for system \eqref{eqkin} taking as conservative terms the ones presented before and as external force field the sum of the two shocks outlined above. Figure \ref{ShockKat} shows the distribution of the income during the time period considered.
		
		Also here, data available for the income distribution at the end of 2005 and the end of 2007 (assuming the final value equal to the mean ones) are marked by asterisks.
		We observe that, during the weeks following the impact of the hurricane, the fractions of the two lowest classes increase, while the fraction of the middle class drastically decreases, and also the middle-upper class fraction shows a decay. This reflects greater mobility of residents to evacuate or move out of the area. On the other hand, in the two years following the event, the higher income fraction tends to increase slightly, showing that wealthy residents are likely to remain and contribute to rebuilding efforts. 
		
\begin{figure}[ht!]
			\centering
			\includegraphics[scale=0.25]{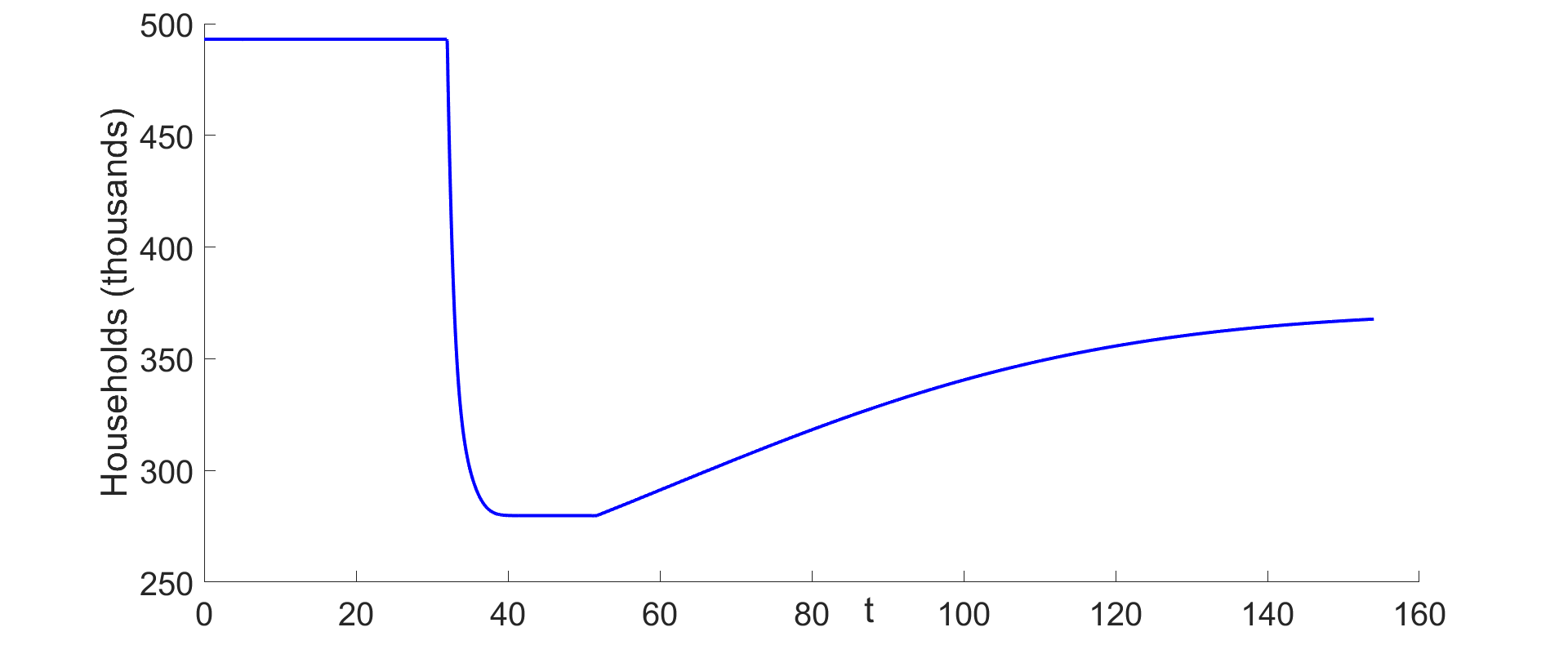}
			\caption{{\it Real scenario  -- impact of the Hurricane Katrina.} 
				Behavior of the total number of households in New Orleans after the drastic reduction due to evacuation immediately after Hurricane Katrina
				and following slow growing in the long-term window representing their return  modeled by system \eqref{eqkin}
				{wi}thout the birth/death terms and taking two consecutive shocks modeled by \eqref{shock4}, with $\nu_1=33, \sigma_1=2,\nu_2=52,\sigma_2=50$, taking as initial data quantity at the beginning of 2005 \cite{shaughnessy2010income}, interaction frequencies as in \eqref{RatesAlKat1} and parameters as in \eqref{ParsAlKat1}-\eqref{ParsAlKat2}.
				{The} time $t$ is in weeks. }
			\label{ShockKatTot}
		\end{figure}

		Moreover, we report in Figure \ref{ShockKatTot} the behavior of the total number of households. We can see how, after the drastic reduction due to evacuation immediately after the catastrophe, in the long-term the total amount still increases slowly,	providing evidence of the fact that, during the months following the storm, the majority of the population, represented by the middle class, was less likely to return.
		
	%%%%%%%%%%%%%%%%
	%%%%%%%%%%%%%%%%
	
	\section{Conclusions and research perspectives}
	\label{secconcl}
	
	In this paper, a nonconservative kinetic framework has been introduced, where both nonconservative binary interactions and external force fields occur. The related Cauchy problem has been investigated, and a first result on existence and uniqueness has been proved in Theorem \ref{thm1}. Then, this framework has been employed, along Section \ref{secmark}, for modeling a closed-market society subjected to shock events. 
	Specific analytical shapes have been introduced to describe the external force field.
	In particular, we have chosen Gaussian functions for slow or sudden shocks,
	essentially because of their mathematical properties and capability of describing abrupt peaks with different intensity
		and width.
	Several scenarios have been investigated, dividing the population into $3$ and $5$ wealth-classes/functional subsystems. 
	
	 {We have presented a first analytical result. Theorem \ref{thm1} provides a local result for the existence and uniqueness of a nonnegative solution $\mathbf{f}(t)$ in a more general framework. Then, we have specialized the model {to the case} when the nonconservative coefficients and the external force field assume a specific analytical form, as required by the application here considered. These choices seem to suggest, as discussed in Remark \ref{rem1}, a possible extension of the previous local result to a global one
	 	%, as discussed in Remark \ref{rem1}, 
	 even if we have not provided a rigorous proof in the more general setting, but rather proposed a first attempt to provide such a global result, having also in mind the recent papers \cite{menale2024kinetic, menale2024nonconservative}, where the external force field is introduced in different contexts with particular analytical shapes.
	 %: in \cite{menale2024kinetic}, a constant external force field is applied in an ecological context, while in \cite{menale2024nonconservative}, although the external force field is not constant in time, it is restricted to a bounded time-interval as the model is used for medical treatment. 
	 We also emphasize that deriving rigorous results would have imposed excessive constraints on the model, potentially shifting the focus away from the specific applications under consideration. However, this decision does not affect the second part of the paper, which focuses on comparing different scenarios by means of dedicated numerical simulations.}
	
	Numerical simulations have confirmed what was empirically expected: the most negative effects of shock events weigh on lower classes of society,  in both cases of $3$ and $5$ income classes. Therefore, the higher the class to which an individual belongs, the more likely he/she will not be a victim of the crisis. Moreover, even when there is more than one shock, for instance, two, the second one impacts heavily the lower classes more, just beaten by the first. This confirms the capability of the model for a more realistic depiction of a closed-market society under the effect of shock events. 
	This particular feature stands out as the primary innovation of this paper, especially from an applied perspective, while the generality of equations \eqref{eqkin}, with time-dependent nonconservative rates and an external force field, provides the newly introduced feature on the theoretical side. 
	
	In addition, Section \ref{secS5} has presented an application of the model to a real-world scenario: the case of Hurricane Katrina. This adaptation demonstrates the model’s ability to describe both the immediate and long-term economic impacts of a large-scale disaster on the population of New Orleans. By incorporating empirical data on income distribution before and after the hurricane, we were able to simulate the effects of sudden and slower shocks on various income classes.
	
	Nevertheless, we are aware that this is only a first attempt at modeling stochastically interacting systems, subjected to nonconservative events, among which slow and sudden shocks {can be considered}. Indeed, a future research perspective is the introduction of external force fields with different analytical shapes for modeling shock events. In this direction, it could be of interest to match models with real data, available in socio-economic literature dealing with shock events. 
	A  {challenging} future research perspective could be the analysis of 
	dependence of the  {solution} on the nonconservative parameters, here considered. {This analysis would pave the way for a more systematic and effective strategy to calibrate the model coefficients to real data. In the present work, this step has represented the initial stage of a broader approach. The ultimate goal is to develop a flexible data-fitting procedure, capable of adapting to a wide range of real-world scenarios.}
	Nevertheless, such a stability analysis would require non-standard tools, 
	since the framework \eqref{eqkin} constitutes a nonautonomous system. 
	{Analogously, this analysis regarding the parameters of the system could help in establishing rigorous global results on existence and uniqueness.}
	
	{Finally, a deep mathematical investigation on the case with sudden shock \eqref{shock2} is mandatory, 
		as the uniqueness of the solution is not generally ensured. In addition, as outlined in this work, the greater the number of wealth classes the population is divided into, the more complex and accurate the model becomes. Therefore, deriving and analyzing a model based on a continuous microscopic variable that accounts for the actual income of each individual rather than the mean income — represents a promising direction for future applications, as recently shown, for instance, in the context of diseases \cite{menale2024nonconservative,conte2024kinetic,oliveira2024reaction,ramos2021kinetic}.}
	
	\bigskip
	
	%%%%%%%%%%%%%%%%
	
	\paragraph*{Acknowledgements.} 
	M.M. and R.T. have written this paper within the activities of GNFM (National Group of Mathematical-Physics) 
	of INDAM (National Institute of High Mathematics). 
	R.T. is a post-doc fellow supported by the National Institute of Advanced Mathematics (INdAM), Italy. 
	The work of R.T. and A.J.S. was carried out in the frame of activities sponsored by the Cost Action CA18232 
	and supported by the {Project ``Mathematical Modelling of Multi-scale Control Systems: applications to human diseases (CoSysM3)'',
	2022.03091.PTDC 
	(\url{https://doi.org/10.54499/2022.03091.PTDC}), Portuguese national funds (OE) through FCT/MCTES.
	A.J.S. thanks the support by Project UID/00013: Centro de Matem\'atica da Universidade do Minho (CMAT/UM).}
	The work of R.T.  was carried out in the frame of activities sponsored by the University of Parma through the action Bando di Ateneo 2022 
	per la ricerca co-funded by MUR-Italian Ministry of Universities and Research - D.M. 737/2021 - PNR - PNRR - NextGenerationEU 
	(project: ``Collective and self-organised dynamics: kinetic and network approaches''). M.M. acknowledges the Italian National Recovery and Resilience Plan (NRRP), M4C2, funded by the European Union – NextGenerationEU (Project IR0000011, CUP B51E22000150006, EBRAINS-Italy), and by the EU Horizon Europe Program under the specific Grant Agreement 101147319, EBRAINS 2.0 project.

	\paragraph*{Conflict of Interest Statement }
	The authors declare that they have no conflict of interest.

	%%%%%%%%%%%%%%%%
	
	\bibliographystyle{unsrt}
	\bibliography{biblio}

\end{document}